\documentclass[12pt,onecolumn,prc,showpacs,preprintnumber,amsmath,
floatfix,amssymb]{revtex4}
\usepackage{epsfig}
\usepackage{graphicx}
\usepackage{dcolumn}
\usepackage{bm}
\def\p{\mbox{\boldmath $p$}}
\def\q{\mbox{\boldmath $q$}}
\def\k{\mbox{\boldmath $k$}}

\allowdisplaybreaks[4]

\begin{document}
\title{Testing of quasi-elastic neutrino charged-current and two-body meson
  exchange current models with the MiniBooNE neutrino data and analysis of these
  processes at energies available at the NOvA experiment}
\author{A.~V.~Butkevich$^{1}$ and S.~V.~Luchuk$^{1,2}$}
\affiliation{$^{1}$Institute for Nuclear Research,
Russian Academy of Sciences, Moscow 117312, Russia\\
$^{2}$Moscow Institute of Physics and Technology, Dolgoprudny 141701, Russia}
\date{\today}
\begin{abstract}

  The charged-current quasi-elastic (CCQE) scattering of muon neutrinos on a
 carbon target is analyzed using the relativistic distorted-wave 
impulse approximation (RDWIA) taking into account the contribution of the 
two-particle and two-hole meson exchange current ($2p-2h$ MEC) to the weak
response functions. A fit the RDWIA+MEC model to the MiniBooNE neutrino data
is performed and the best fit value of nucleon axial mass $M_A=1.2 \pm 0.06$
GeV is obtained. We also extract the values of the axial form factor
$F_A(Q^2)$ as a function of the squared momentum transfer $Q^2$ from
the measured $d\sigma/dQ^2$ cross section. The flux-integrated CCQE-like
differential cross sections for neutrino scattering at energies of the NOvA
experiment are estimated within the RDWIA+MEC approach .
  
\end{abstract}

\maketitle

\section{Introduction}

The high-intensity muon-(anti)neutrino beams used in long-baseline neutrino
oscillation experiments are peaked in the energy range from a few hundreds of
MeV to several GeV. In this energy regime the dominant contribution to
neutrino-nucleus scattering comes from the charged-current quasielastic (CCQE)
interaction, two-body meson exchange current (MEC), and resonance production
processes. To determine values of neutrino oscillation parameters, the
probabilities of $\nu_{\mu}$ disappearance and $\nu_e$ appearance versus
neutrino energy are measured in these experiments. The accuracy of these
measurements depends explicitly on how well we are able to evaluate the energy
of the incoming neutrino.  This energy can be estimated from the lepton and
hadron energies visible in the final state after the neutrino has interacted.
Thus the total hadronic deposit is the necessary piece of information for the
calorimetric method.

Because the CCQE is a two-body process, the incoming
neutrino energy can be calculated using only outgoing lepton kinematics. The
measurement of muon momentum and angle allows the estimation of the neutrino
energy $\varepsilon_{\nu}$ and the squared four-momentum transfer $Q^2$. This
reconstruction method (kinematic method) works well if the true nature of
events were indeed a CCQE process. Nuclear effects, the final state interaction
(FSI) between of the outgoing particles and the residual nucleus as well as
interactions which are not distinguishable from CCQE in the final state --
bias or smear the reconstructed neutrino energy. Therefore a good understanding
 of these effects is critical. 

To model the CCQE scattering from a nuclei, most event generators
are based on the relativistic Fermi gas model (RFGM)~\cite{Moniz}. In this model
the nucleus is described as a system of quasi-free nucleons with a flat momentum
distribution up to the same Fermi momentum $p_F$ and nuclear binding energy
$\epsilon_b$. With the assumption of the conserved vector current, the only
parameter of the weak current which is not well constrained by electron
scattering data is the axial nucleon form factor $F_A(Q^2)$. In most analysis of
the CCQE interaction, the dipole parametrization of $F_A(Q^2)$ with one
parameter, the axial mass $M_A$ is used. Note that dipole parametrization has
no strict theoretical basis, and the choice of this parametrization is made by
analogy with electromagnetic form factors.

The value of $M_A$ is obtained from a fit to observed $Q^2$ distribution of
events, differential, and total (anti)neutrino  CCQE cross sections. Results
from global analysis of neutrino-deuterium scattering experiments are very
widely spread and the formal averaging of $M_A$ values was done in
Ref.~\cite{Bern}: $M_A=1.026 \pm 0.021$ GeV. This result is also known as the
world-averaged value of the axial mass. The NOMAD experiment has reported
result on neutrino CCQE scattering on carbon: $M_A=1.05 \pm 0.02 \pm 0.06$ GeV~
\cite{NOMAD}. The MINERvA experiment~\cite{Min1, Min2} has shown good
agreement within the RFGM with $M_A\approx 1$ GeV, but requires an enhancement
to the transverse response function. A recent reanalysis of the MINERvA
flux~\cite{Leo} results in the increases to the normalization of previous
cross sections~\cite{Beta, Min3} and invalidates conclusions from
Refs.~\cite{Min1, Min2}.

On the other hand the differential cross sections measured by the MiniBooNE
collaboration~\cite{MiniB1, MiniB2, MiniB3} can be described within the RFGM
only with large value of $M_A=1.35 \pm 0.017$ GeV. The absolute values of the
differential and total cross sections are about 30\% larger compared to
NOMAD results. Large values of axial mass $M_A\approx 1.1 - 1.3$ GeV
have also obtained in other experiments using heavy nuclear targets~
\cite{K2K1, K2K2, MINOS,T2K1}. 

These results have stimulated many theoretical studies trying to explain the
apparent discrepancy between the data and theoretical predictions, and present a
considerable challenge to neutrino oscillation experiments. A wide variety of
models has been proposed to describe CCQE-like cross sections, identified
experimentally as processes in which only a final charged lepton with
multinucleon excitations is detected, but the pion absorption contribution is
subtracted.  The data without subtracting any intrinsic background is called
CC0$\pi$. A review of the available CCQE-like cross section models can be found
in Refs.~\cite{Garv, Katori, Alvar}.

Based on the results from different groups it is shown that CCQE-like data
are really a combination of the genuine QE and of the two-particle and two-hole
meson exchange current ($2p-2h$ MEC) contributions to weak response functions.
Such excitations are induced by two-body currents, hence they go beyond
the usual impulse approximation (IA) scheme, in which the probe interacts with
only single nucleon and corresponds to the $1p-1h$ excitations. To describe the
genuine QE a model should in principle include the nuclear mean field and
nucleon-nucleon ($NN$) short and long-range correlations in the ground state, as
well as final state interaction of the outgoing nucleon with the residual
nucleus. More sophisticated descriptions of the CCQE interaction that the RFGM
provides are available from a number authors~\cite{
  Maier, Ryck1, Meucci, BAV1, BAV2, BAV3, Martini1, Ryck2, Nieves1, Benhar,
  Amaro1, Mosel1}. Note that there exist some differences already at level of
the genuine quasielastic scattering.

The transverse enhancement effective model to account for MEC effects has been
proposed in Ref.~\cite{Bodek}. In this model the magnetic form factor for
nucleon bound in carbon are modified to describe the enhancement in the
electron-carbon QE cross section. An enhancement of the axial nucleon mass in
the nonrelativistic continuum random phase approximation has been regarded in
Ref.~\cite{Ryc2}. The contribution of $np-nh$ channel is also taken into
account through phenomenological approach in Ref.~\cite{Mosel2}.

The most complete theoretical calculations of $2p-2h$ cross sections are
performed by different groups~
\cite{Martini2, Martini3, Nieves2, Nieves3, Pace, Amaro2, Amaro3,
  Amaro4}. In Refs.~\cite{Martini2, Martini3, Nieves2, Nieves3} the models start
from a local Fermi gas picture of the nucleus and take into account long-range
random phase approximation (RPA) corrections, but ignore the shell structure
of nucleus and FSI effects. In the $2p-2h$ sector both models use the Fermi
gas approximation. The short-range $NN$-correlations are included by
considering an additional two-body correlation current. As result, the
$NN$-correlations and $NN$-correlations-MEC-interference naturally apper
(RPA-MEC approach).

In the SuSA approach~\cite{Pace, Amaro1, Amaro2, Amaro3, Amaro4} a
superscaling analysis of electron scattering result is used to calculate
neutrino cross sections. The effects of the short-range $NN$-correlations in
the $1p-1h$ sector are effectively included via the superscaling function.
In Ref.~\cite{Amaro4} the SuSAv2 model is combined with MECs in $2p-2h$ sector
by using accurate parametrizations of the exact calculation of electroweak MEC
response functions~\cite{Pace, Amaro2} (SuSAv2-MEC approach). The
$NN$-correlations and $NN$-correlations-MEC-interference are absent in the
$2p-2h$ MEC contributions.

Another approach which goes beyond the impulse approximation was
developed in Ref.~\cite{BAV4}. In this work a joint calculation of the CCQE
and $2p-2h$ contributions to the lepton scattering cross sections on carbon,
using relativistic distorted-wave impulse approximation (RDWIA) for quasielastic
response functions in the electroweak sector (RDWIA+MEC approach) was performed.
The RDWIA, which takes into account the nuclear shell structure and final-state
interaction effects, was developed to describe of the QE electron-nucleus
scattering ~\cite{Kelly1, Fissum, Kelly2}. Results of the analyze data for
${}^{12}$C$(e,e'p)$ based on upon the RDWIA can be found in Refs.~\cite{Kelly2,
  BAV2}, which show that the nucleon momentum distributions are described well
by mean-field calculations. From this analysis it follows that fragmentation of
the $1p$ strength in carbon by collective modes is largerly confined to
excitation energy be low 10 MeV and approximately 84\% of the independent
particle shell model (IPSM) is found in the missing energy bin 15-25 MeV. In
our approach~\cite{BAV1, BAV2} the effects of the short-range
$NN$-correlations, leading to appearance of a high-momentum and high-energy
distribution in the target are estimated.

We explicitly added the MEC contributions (without the $NN$-correlations-MEC-
interference) to the genuine QE interaction, as in the SuSA-MEC approach~
\cite{Amaro4}. The functional forms employed for the parametrizations of the
MEC transverse electromagnetic vector response, and for the axial and vector
components of the weak response were detailed in Refs.~\cite{Amaro3, Amaro4}.
The RDWIA+MEC approach was successfully tested against ${}^{12}$C$(e,e')$ data~
\cite{BAV4}. 

Although theoretical calculations of the CCQE-like neutrino-nucleus cross
sections have been performed by many groups using different approaches, at this
moment there is no progress in a quantitative description of the data and it
is not clear which models fit the global data best. For example, the global
fit performed in Ref.~\cite{Wilk} shows very poor results. One of the goals of
this paper is to fit the RDWIA+MEC model to the MiniBooNE data~\cite{MiniB1}
for neutrino scattering off carbon. Within this approach we extract the value
of the axial mass from measured flux-integrated $d\sigma/Q^2$ and
$d^2\sigma/dTd\cos\theta$ ($T$ and $\theta$ are, correspondingly, kinetic energy
and muon scattering angle) differential cross sections. In addition we
determine the values of the axial form factor $F_A(Q^2)$ as a function of $Q^2$,
using the method described in Ref.~\cite{BAV5}. Previously our
constraint on the $M_A\approx 1.37$ GeV was obtained within the
RDWIA~\cite{BAV5, BAV6}. This work improves the previous situation by
including $2p-2h$ MEC contributions. A second topic addressed in this paper is
calculations of the CCQE-like flux-integrated differential and
double-differential cross sections at energies of the NOvA experiment~
\cite{NOvA1, NOvA2}. We evaluated these cross sections within the RDWIA+MEC
approach with value of $M_A$ extracted from the MiniBooNE data.

This article is organized as follows. In Sec. II we briefly present the
RDWIA+MEC model and the procedure which allows the determination of values of
the axial form factor from the $d\sigma/dQ^2$ differential cross section.
Section III presents results of this model to the MiniBooNE data, extraction
of the $F_A(Q^2)$, and calculations of the flux-integrated differential cross
sections for the NOvA experiment. Our conclusions are summarized in Sec. IV.

\section{Formalism of quasielastic scattering, RDWIA, $2p-2h$ MEC responses,
  and flux-integrated cross sections.}

We consider neutrino charged-current QE inclusive
\begin{equation}\label{qe:incl}
\nu_{\mu}(k_i) + A(p_A)  \rightarrow \mu(k_f) + X                      
\end{equation}
scattering off nuclei in the one-W-boson exchange approximation. Here 
$k_i=(\varepsilon_i,\k_i)$ and $k_f=(\varepsilon_f,\k_f)$ are the initial and 
final lepton momenta, $p_A=(\varepsilon_A,\p_A)$ is the initial target momenta,
 $q=(\omega,\q)$ is the momentum transfer carried by 
the virtual W-boson, and $Q^2=-q^2=\q^2-\omega^2$ is the W-boson virtuality. 

\subsection{CCQE-like quasielastic lepton-nucleus cross sections}

In the inclusive reactions (\ref{qe:incl}) only the outgoing lepton is
detected and the differential cross sections can be written as
\begin{eqnarray}
\frac{d^3\sigma}{d\varepsilon_f d\Omega_f } =
\frac{1}{(2\pi)^2}\frac{\vert\k_f\vert}                            
{\varepsilon_i} \frac{G^2\cos^2\theta_c}{2} L_{\mu \nu}
{W}^{\mu \nu},
\label{CS}
\end{eqnarray}
where $\Omega_f=(\theta,\phi)$ is the solid angle for the muon momentum,  
$G \simeq$ 1.16639 $\times 10^{-11}$ MeV$^{-2}$ is
the Fermi constant, $\theta_C$ is the Cabbibo angle
($\cos \theta_C \approx$ 0.9749), $L_{\mu \nu}$ is the lepton tensor,
and $W^{\mu \nu}$ are the weak CC nuclear tensors. In terms of response
functions the cross sections reduce to
\begin{eqnarray}
\frac{d^3\sigma}{d\varepsilon_f d\Omega_f}=
\frac{G^2\cos^2\theta_c}{(2\pi)^2} \varepsilon_f
\vert \k_f \vert\big ( v_0R_0 + v_TR_T                            
+ v_{zz}R_{zz} -v_{0z}R_{0z}- hv_{xy}R_{xy}\big),
\label{CSR}
\end{eqnarray}
where the coupling coefficients $v_k$ , whose expressions are given
in~\cite{BAV1} are kinematic factors depending on the lepton's kinematics.
The response functions are given in terms of components of the hadronic tensors
\begin{subequations}
\begin{align}
R_0 & = W^{00},\\
R_T & = W^{xx} + W^{yy},\\
R_{0z}&  = W^{0z} + W^{z0},\\
R_{zz} & = W^{zz}, \\                                             
R_{xy} & =  i\left(W^{xy}-W^{yx}\right) 
\end{align}
\label{R}
\end{subequations}
and depend on the variables ($Q^2, \omega$) or ($|q|,\omega$).   

All the nuclear structure information and final state interaction effects are
contained in the weak CC nuclear tensor. It 
is given by the bilinear products of the transition matrix 
elements of the nuclear CC operator $J^{cc}_{\mu}$ between the initial nucleus
state $|A\rangle$ and the final state $|X_f\rangle$ as 
\begin{eqnarray}
W_{\mu \nu } &=& \sum_f \langle X_f\vert                           
J^{(cc)}_{\mu}\vert A\rangle \langle A\vert
J^{(cc)\dagger}_{\nu}\vert X_f\rangle,              
\label{W}
\end{eqnarray}
where the sum is taken over undetected states $X_f$. This equation includes all
possible final states. Thus, the hadron tensor can be expanded as the sum of
the $1p-1h$ and $2p-2h$, plus additional channels: 
\begin{eqnarray}
W^{\mu \nu } &=& W^{\mu \nu}_{1p1h} + W^{\mu \nu}_{2p2h} + \cdots,      
\label{W_12}
\end{eqnarray}
where the $1p-1h$ channel gives the CCQE response functions and the $2p-2h$
hadronic tensor determines the $2p-2h$ MEC response functions. Thus, the
functions $R_i$~(\ref{R}) can be written as a sum of the CCQE ($R_{i,QE}$) and
MEC ($R_{i,MEC}$) response functions
\begin{eqnarray}
R_i &=& R_{i,QE} + R_{i,MEC}.                                      
\label{R_12}
\end{eqnarray}

\subsection{RDWIA model}

We describe genuine CCQE neutrino-nuclear scattering in the impulse 
approximation, assuming that the incoming neutrino interacts with only 
one nucleon, which is subsequently emitted, while the remaining (A-1) nucleons 
in the target are spectators. The nuclear current is written as the sum of 
single-nucleon currents.

The single-nucleon charged current has $V{-}A$ structure
$J^{\mu} = J^{\mu}_V +J^{\mu}_A$. For a free-nucleon vertex function
$\Gamma^{\mu} = \Gamma^{\mu}_V + \Gamma^{\mu}_A$ we use the CC2 vector 
current vertex function
$\Gamma^{\mu}_V = F_V(Q^2)\gamma^{\mu} + {i}F_M(Q^2)\sigma^{\mu \nu}q_{\nu}/2m$, 
where $\sigma^{\mu \nu}=i[\gamma^{\mu},\gamma^{\nu}]/2$, $F_V$ and
$F_M$ are the weak vector form factors. They are related to the
corresponding electromagnetic ones for proton and neutron by the hypothesis of
the conserved vector current. We use the approximation of Ref.~\cite{MMD} for
the vector nucleon form factors. Because the bound nucleons are
off-shell we employ the de Forest prescription~\cite{deFor} and use the Coulomb
 gauge for the off-shell vector current vertex $\Gamma^{\mu}_V$.

The axial current vertex function can be written in terms of the axial
$F_A(Q^2)$ and pseudoscalar $F_P$ form factors
\begin{equation}
\Gamma^{\mu}_A = F_A(Q^2)\gamma^{\mu}\gamma_5 + F_P(Q^2)q^{\mu}\gamma_5.      
\label{Eq8}
\end{equation}
These form factors are parameterized using a dipole approximation: 
\begin{equation}
F_A(Q^2)=\frac{F_A(0)}{(1+Q^2/M_A^2)^2},\quad                              
F_P(Q^2)=F_A(Q^2)F'_P(Q^2),
\label{Eq9}
\end{equation}
where $F'_P=2m^2/(m_{\pi}^2+Q^2), F_A(0)=1.267$, $M_A$ is the axial mass, and
$m_\pi$ is the pion mass. Then the axial current vertex function can be written
 in the form
\begin{eqnarray}
\label{Eq10}
\Gamma_A^{\mu} &=& F_A(Q^2)[\gamma^{\mu}\gamma_5 +F'_P(Q^2)q^{\mu}\gamma_5]  
\end{eqnarray}
and the axial vector current can be factorized as
\begin{eqnarray}
\label{Eq11}
J_A = F_A(Q^2)J'_A(Q^2),                                      
\end{eqnarray}
where $J'_A = \gamma^{\mu}\gamma_5 +F'_P(Q^2)q^{\mu}\gamma_5$  

In the RDWIA, the relativistic wave functions of the bound nucleon states are 
calculated in the IPSM as the self-consistent 
solutions of a Dirac equation, derived within a relativistic mean field 
approach, from a Lagrangian containing $\sigma, \omega$, and $\rho$ mesons 
(the $\sigma-\omega$ model)\cite{Serot,Horow}. According to the JLab
data~\cite{Dutta, Kelly2} the occupancy of the independent particle shell model
orbitals of ${}^{12}$C equals on average 89\%. In this work, we assume that the
missing strength (11\%) can be attributed to the short-range $NN$-correlations
 in the ground state, leading to the appearance of the high-momentum and 
high-energy component in the nucleon distribution in the target.  
In the RDWIA, the final state interaction effects for the outgoing nucleon are 
taken into account. The distorted-wave function of the knocked out nucleon is 
evaluated as a solution of a Dirac equation containing a phenomenological 
relativistic optical potential. 
The EDAD1 parametrization~\cite{Cooper} of the relativistic optical potential
for carbon was used in this work. We calculated the inclusive cross sections
with the EDAD1 relativistic optical potential in which only the real part was
included.

The cross sections with the FSI effects in the presence of the 
short-range $NN$-correlations were calculated by using the method proposed in
 Ref.~\cite{BAV1} with the nucleon high-momentum distribution from 
Ref.~\cite{Atti} that was renormalized to value of 11\%. In this approach, the 
contribution of the $NN$-correlated pairs is evaluated in impulse 
approximation, {\it i.e.}, the virtual W-boson couples to only one member 
of the $NN$-pair. It is a one-body process that leads to the emission of two 
nucleons ($2p-2h$ excitation). 

\subsection{$2p-2h$ excitation}

In the present work we evaluate the weak MEC response functions $R_{i,MEC}$ of
neutrino scattering on carbon, using accurate parametrizations of the exact MEC
calculation~\cite{Amaro2}. In order to evaluate the $2p-2h$ hadronic tensor
$W^{\mu \nu}_{2p 2h}$, in Ref.~\cite{Amaro2} the RFGM was chosen to describe
the nuclear ground state. The short-range $NN$-correlations and FSI effects
were not considered in this approach. The elementary hadronic tensor is given
by bilinear product of the matrix elements of the two-body weak 
(containing vector and axial components) MEC. Only one-pion exchange is 
included. 

The two-body current operator is obtained from the pion 
production amplitudes for the nucleon while coupling a second 
nucleon to the emitted pion. The resulting MEC operator can be written as a
sum of seagull, pion-in-flight, pion-pole, and Delta-pole operators. The
$\Delta$-peak is the main contribution to the pion production cross section and 
the MEC peak is located in the ``dip'' region between the QE and 
Delta peaks.

The functional forms employed for these parametrizations as functions of
$(\omega, |\q|)$  are  valid in the range of momentum transfer
$|\q|=200 \div 2000$ MeV. The expressions for the fitting parameters are
described in detail in Refs.~\cite{Amaro3, Amaro4, Megias}. Results of
lepton-nucleus cross sections obtained using these MEC parametrizations were
successfully tested against the experimental world data for
${}^{12}$C~\cite{Amaro3, Megias2, BAV4}. 

\subsection{Flux integrated cross sections and the method for extraction of
  $F_A(Q^2)$ from $d\sigma/dQ^2$ distribution. }

The inclusive weak hadronic tensor is bilinear in the transition matrix
elements of the nuclear weak current operators
$W_{\mu\nu} = \langle J_{\mu} J^{\dagger}_{\nu}\rangle $, where the angle 
brackets denote products of matrix elements appropriately averaged over 
initial states and summed over final states. By using Eq.~\eqref{Eq11} the
axial vector current can be written as $J = J_V +F_AJ'_A$. The expressions for
the inclusive CCQE cross sections in terms of vector $\sigma^V$, axial
$\sigma^A$, and vector-axial $\sigma^{VA}$ cross sections then given
by~\cite{BAV5}
\begin{subequations}
\begin{align}
\label{Eq12}
\frac{d\sigma^{\nu}}{dQ^2}(Q^2,\varepsilon_i) &= 
\sigma^V(Q^2,\varepsilon_i) + F^2_A(Q^2)\sigma^A(Q^2,\varepsilon_i) + 
hF_A(Q^2)\sigma^{VA}(Q^2,\varepsilon_i)\\                                 
\frac{d^2\sigma^{\nu}}{dTd\cos\theta}(T,\cos\theta,\varepsilon_i) &=
\sigma^V(T,\cos\theta,\varepsilon_i) + F^2_A(Q^2)\sigma^A
(T,\cos\theta,\varepsilon_i)
\notag \\
& + hF_A(Q^2)\sigma^{VA}(T,\cos\theta,\varepsilon_i),
\end{align}
\end{subequations}
where $\sigma^V$ is the cross section $d\sigma/dQ^2(d^2\sigma/dTd\cos\theta)$
calculated with $F_A=0$ and $\sigma^A$ is the cross section
$d\sigma/dQ^2(d^2\sigma/dTd\cos\theta)$ calculated with $F_V=F_M=0, F_A=1$.
The vector-axial cross section $\sigma^{VA}$, arising from the interference
between the vector and axial currents can be written as
\begin{eqnarray}
\label{Eq13}
\sigma^{VA} &=& h[\sigma(F_A=1) - \sigma^V - \sigma^A],               
\end{eqnarray}
where $\sigma(F_A=1)$ is the $d\sigma/dQ^2 (d^2\sigma/dTd\cos\theta)$ cross
section, calculated with $F_A(Q^2)$=1.

The flux integrated cross section can be written as a sum of the flux
integrated QE and $2p-2h$ MEC contributions
\begin{eqnarray}
\label{Eq14}
\left\langle \frac{d\sigma}{dQ^2}(Q^2)\right\rangle =  
\left\langle \frac{d\sigma_{QE}}{dQ^2}(Q^2)\right\rangle +
\left\langle \frac{d\sigma_{MEC}}{dQ^2}(Q^2)\right\rangle,             
\end{eqnarray}
where
\begin{eqnarray}
\label{Eq15}
\left\langle \frac{d\sigma_j}{dQ^2}(Q^2)\right\rangle &=&              
\int D_{\nu}(\varepsilon_i)
\frac{d\sigma_j}{dQ^2}(Q^2,\varepsilon_i) 
d\varepsilon_i,
\end{eqnarray}
and $j=QE,MEC$. The weight functions $D_{\nu}$ are defined as 
\begin{eqnarray}
\label{Eq16}
D_{\nu}(\varepsilon_i) &=& I_{\nu}(\varepsilon_i)/                    
\Phi_{\nu},
\end{eqnarray}
where $I_{\nu}(\varepsilon_i)$ is the neutrino spectrum and $\Phi_{\nu}$ is the
 integral neutrino flux.

 The value of $F_A(Q^2)$ as a function of $Q^2$ can be extracted as the solution
 to the equation
\begin{eqnarray}
\label{Eq17}
\left\langle \frac{d\sigma_{QE}}{dQ^2}(Q^2)\right\rangle =  
\left\langle \frac{d\sigma}{dQ^2}(Q^2)\right\rangle -               
\left\langle \frac{d\sigma_{MEC}}{dQ^2}(Q^2)\right\rangle, 
\end{eqnarray}
using the neutrino CCQE-like scattering data for
$\langle d\sigma^{\nu}/dQ^2\rangle$ and 
\begin{eqnarray}
\label{Eq18}
\left\langle \frac{d\sigma_{QE}}{dQ^2}(Q^2)\right\rangle = & 
\langle\sigma^V(Q^2)\rangle +                                       
F^2_A(Q^2)\langle\sigma^A(Q^2)\rangle
 + h F_A(Q^2)\langle\sigma^{VA}(Q^2)\rangle,
\end{eqnarray}
where ${\left\langle \sigma^j(Q^2)\right\rangle}$ are the flux-integrated
vector, axial and vector-axial $(j=V,A,VA)$ cross sections. Note,
that the values of the axial form factor extracted from the CCQE-like cross
sections are model dependent and implicitly include the uncertainties in
the $F_V$, $F_M$, $\nu_{\mu}$-flux, and $2p-2h$ MEC contributions.

\section{Results and analysis}

Our main interest is to show the capability of the present model, RDWIA+MEC,
to describe successfully the MiniBooNE neutrino scattering data and calculate
within this approach the CCQE-like flux-integrated cross sections at energies
available at the NOvA experiment.

\subsection {Fit of the RDWIA+MEC model to the neutrino MiniBooNE data}

The MiniBooNE neutrino data set~\cite{MiniB1}, obtained in a kinematic range
that significantly overlaps with the range available to the
NOvA experiment, is used in the CCQE-like fit. These data
have been released as flux-integrated double-differential cross section
$d^2\sigma/dTd\cos\theta$ and as differential cross section $d\sigma/dQ^2$  in
the range $0<Q^2<2$ (GeV/c)$^2$. The data release included the diagonal elements
of the shape-only covariant matrix for each bin and correlations between bins
were not presented. The normalization uncertainly was given as 10.7\%. In our
analysis we use the CCQE corrected sample with purity 77\%. In the
MiniBooNE antineutrino data set~\cite{MiniB3} the correction algorithm for the
antineutrino data is more complicated then for neutrino mode sample, due to
the relatively high $\nu_{\mu}$  contamination in the $\bar{\nu}_{\mu}$
beam. There is also a large CC1$\pi^{-}$ background in the $\bar{\nu}_{\mu}$
CCQE sample, as most of the $\pi^{-}$ are absorbed. As a result of the two
large backgrounds in the antineutrino sample, the purity of the antineutrino
CCQE-like sample is 61\%.
\begin{figure*}
  \begin{center}
    \includegraphics[height=10cm,width=12cm]{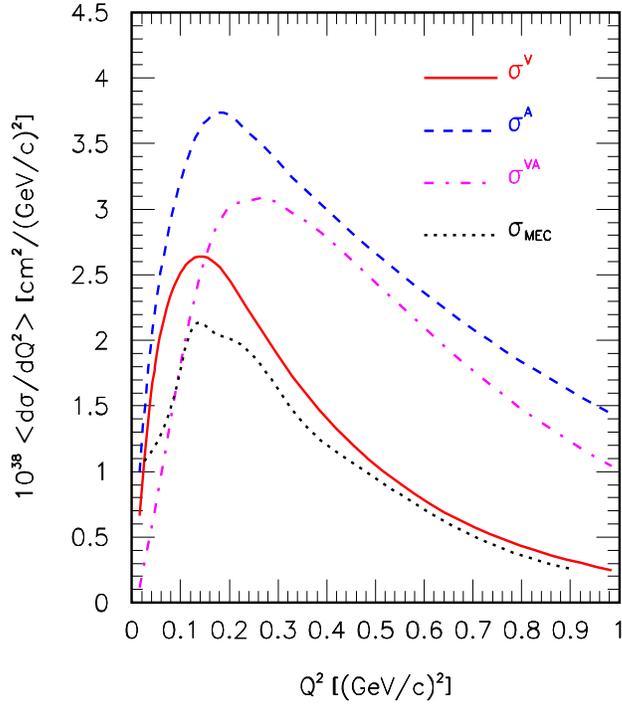}
  \end{center}
  \caption{\label{Fig1} Flux-integrated $\langle \sigma^{V} \rangle$
      (solid line), $\langle \sigma^{A} \rangle$ (dashed line),
      $\langle \sigma^{VA} \rangle$ (dashed-dotted line), and
    $\langle d\sigma_{MEC}/dQ^2 \rangle$ (dotted line) cross sections of
    $\nu_{\mu}$ scattering on ${}^{12}$C as functions of $Q^2$.}
\end{figure*}

\begin{figure*}
  \begin{center}
    \includegraphics[height=14cm,width=14cm]{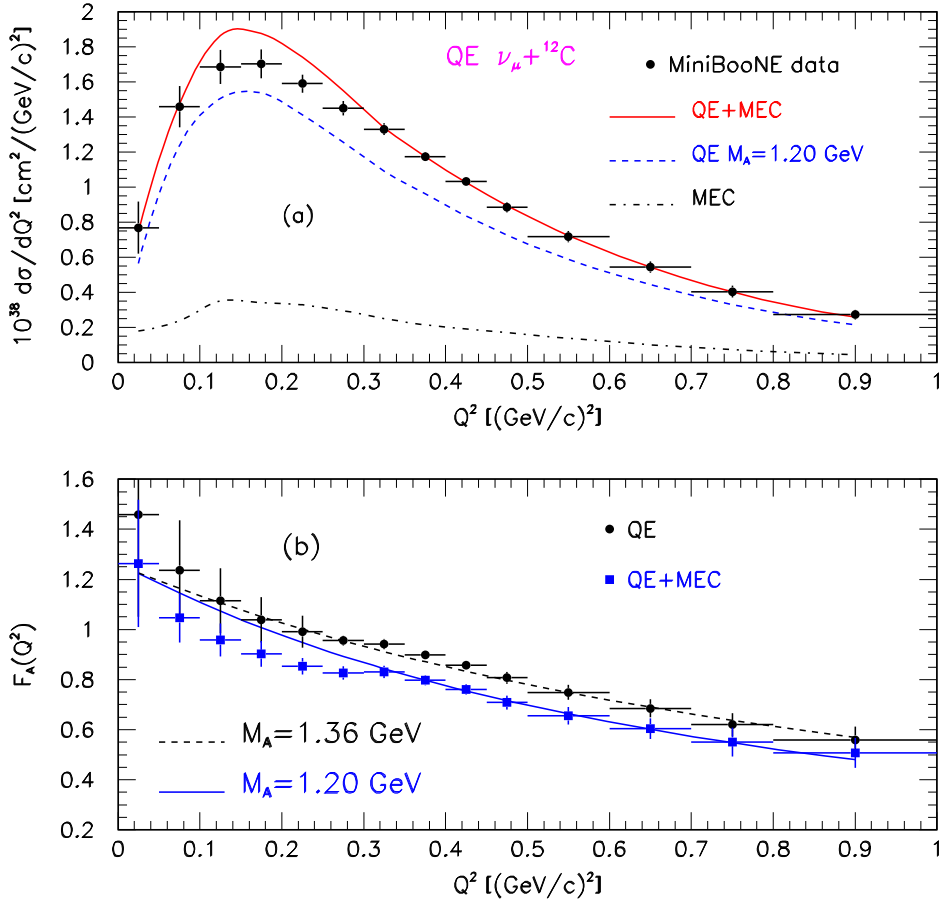}
  \end{center}
  \caption{\label{Fig2}Flux-integrated $d\sigma/dQ^2$ cross section per target
    neutron for the $\nu_{\mu}$ CCQE-like scattering (upper panel) and axial
    form factor $F_A(Q^2)$ extracted from the MiniBooNE data as functions of
    $Q^2$. Upper panel: Calculations from the RDWIA+MEC (solid line), RDWIA
    (dashed line) with $M_A=1.2$ GeV, and $2p-2h$ MEC contributions
    (dashed-dotted line). Lower panel: Filled squares (filled circles) are the
    axial form factor extracted within the RDWIA+MEC (RDWIA), and the solid
    (dashed) line is the result of the dipole parametrization with
    $M_A=1.2(1.36)$ GeV.} 
\end{figure*}

To extract the values of the axial form factor $F_A(Q^2)$ as a function of
$Q^2$, using the measured neutrino flux-integrated
$\langle d\sigma/dQ^2 \rangle$ cross section, we calculated the
$\langle \sigma^{V} \rangle, \langle \sigma^{A} \rangle, \langle \sigma^{VA}
\rangle $, and $\langle d\sigma_{MEC}/dQ^2 \rangle$ cross
sections with the booster neutrino beam line $\nu_{\mu}$ flux~\cite{MiniB1}. In
Fig.~\ref{Fig1} these cross sections for $\nu_{\mu}$ scattering on ${}^{12}$C are
shown against $Q^2$. In Fig.~\ref{Fig2}(a) we show the measured flux-integrated
$\langle d\sigma/dQ^2 \rangle $ as a function of $Q^2$.
To extract the values of
$F_A$ this cross section with the shape-only error was used in Eq.~\eqref{Eq17}.
The results, $F_A(Q^2)$ as a function of $Q^2$, are shown in
Fig.~\ref{Fig2}(b). Also shown
in this figure are the results from Ref.~\cite{BAV5}, obtained within the RDWIA,
{\it i.e.} without the $2p-2h$ MEC contributions. The axial form factor values
extracted in the RDWIA approach agree well with the dipole parametrization with
$M_A=1.36$ GeV. As observed, results in Fig.~\ref{Fig2}(b) clearly show the
relevant role played by the $2p-2h$ MEC contributions.
\begin{figure*}
  \begin{center}
    \includegraphics[height=14cm,width=14cm]{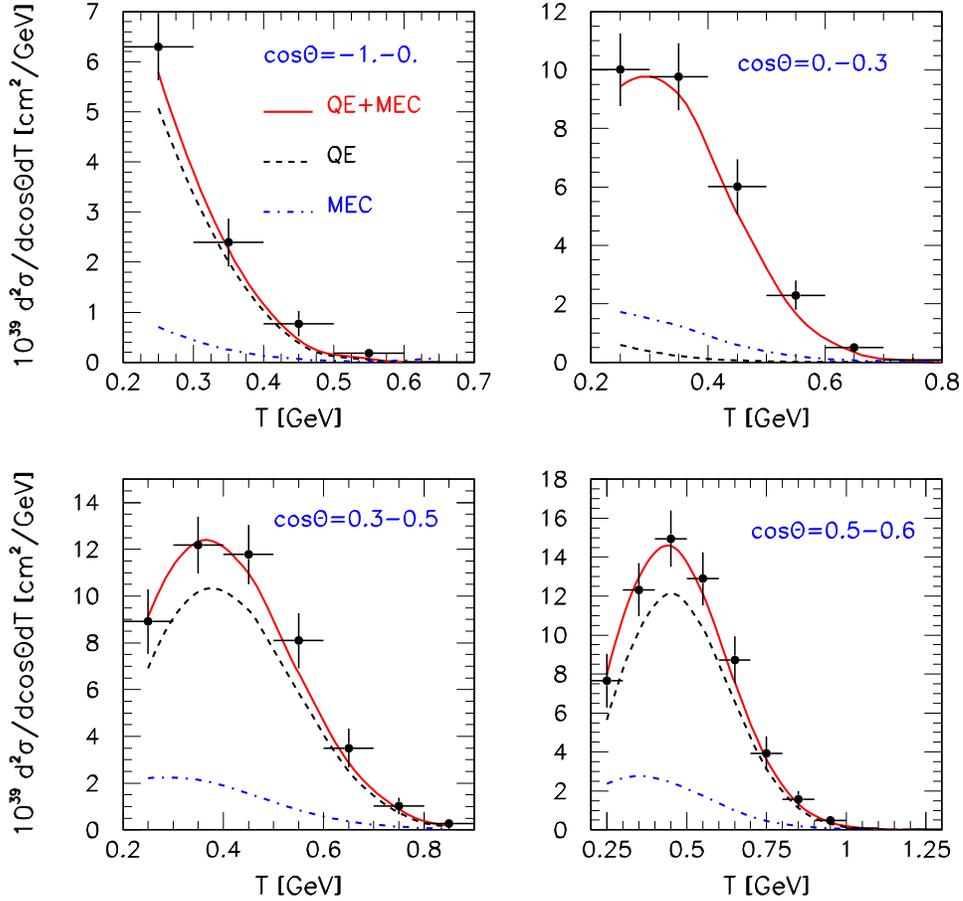}
  \end{center}
  \caption{\label{Fig3}Flux-integrated $d^2\sigma/dTd\cos\theta$ cross section
    per target neutron for the $\nu_{\mu}$ CCQE-like scattering as a function of
    muon kinetic energy for the four muon scattering angle bins:
    $\cos\theta=$(-1-0), (0-0.3), (0.3-0.5), and (0.5 - 0.6). As shown in the
    key, cross sections were calculated within the RDWIA+MEC approach. The
    QE and $2p-2h$ MEC contributions are also presented separately. The
    MiniBooNE data are shown as points with the shape error only.} 
\end{figure*}
\begin{figure*}
  \begin{center}
    \includegraphics[height=14cm,width=14cm]{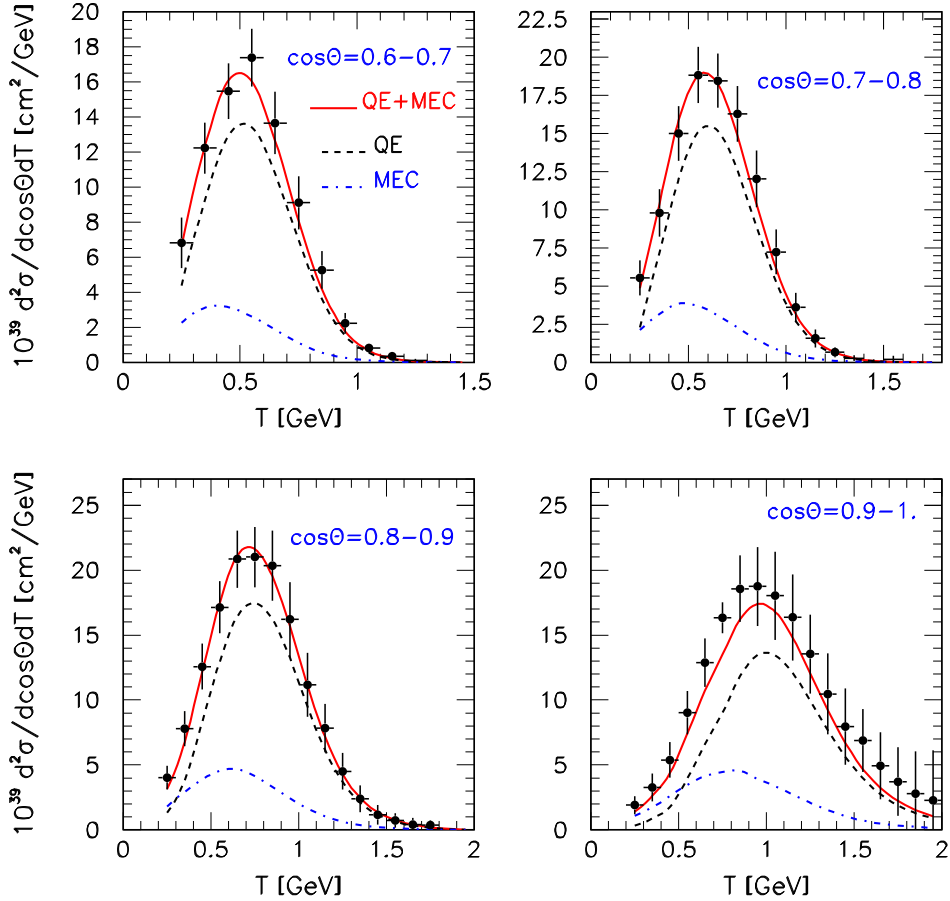}
  \end{center}
  \caption{\label{Fig4} Same as Fig.~\ref{Fig3} but for muon scattering angle
    bins:
$\cos\theta=$(0.6-0.7), (0.7-0.8), (0.8-0.9), and (0.9-1).} 
\end{figure*}
The values of $F_A$
obtained with the RDWIA are higher and decrease with $Q^2$ more
slowly than corresponding values extracted within the RDWIA+MEC approach.   

We also performed a shape-only fit the RDWIA+MEC model to the MiniBooNE
neutrino data with only the axial mass $M_A$ as a variable model parameter.
Ref.~\cite{Wilk2} shows that the best fit parameter values are not
significantly altered by including the MiniBooNE normalization uncertainties in
the CCQE fit. The fits were made to the single-differential $d\sigma/dQ^2$
(1D fit), double-differential $d^2\sigma/dTd\cos\theta$ (2D fit) cross
sections, and their combination (1D+2D main fit), using the $\chi^2$ statistic
\begin{align}
  \chi^{2} &= \sum^{N}_{k=1}\left\lbrack \frac{(d\sigma/dQ^2)_{k}^{data}-
    (d\sigma/dQ^2)_{k}^{th}}{\Delta(d\sigma/dQ^2)_{k}} \right \rbrack^{2} 
\rightarrow \mathrm{1D} \notag\\
&+ \sum^{M}_{l=1}\left\lbrack\frac{(d^2\sigma/dTd\cos\theta)_{l}^{data}-
  (d^2\sigma/dTd\cos\theta)_{l}^{th}}{\Delta(d^2\sigma/dTd\cos\theta)_{l}} \right
\rbrack^{2}\rightarrow \mathrm{2D},
\end{align}
where $\Delta(d\sigma/dQ^2)_{k}$ and $\Delta(d^2\sigma/dTd\cos\theta)_{l}$ are
the diagonals of the MiniBooNE shape-only covariance matrices for neutrino
results.
\begin{figure*}
  \begin{center}
    \includegraphics[height=14cm,width=14cm]{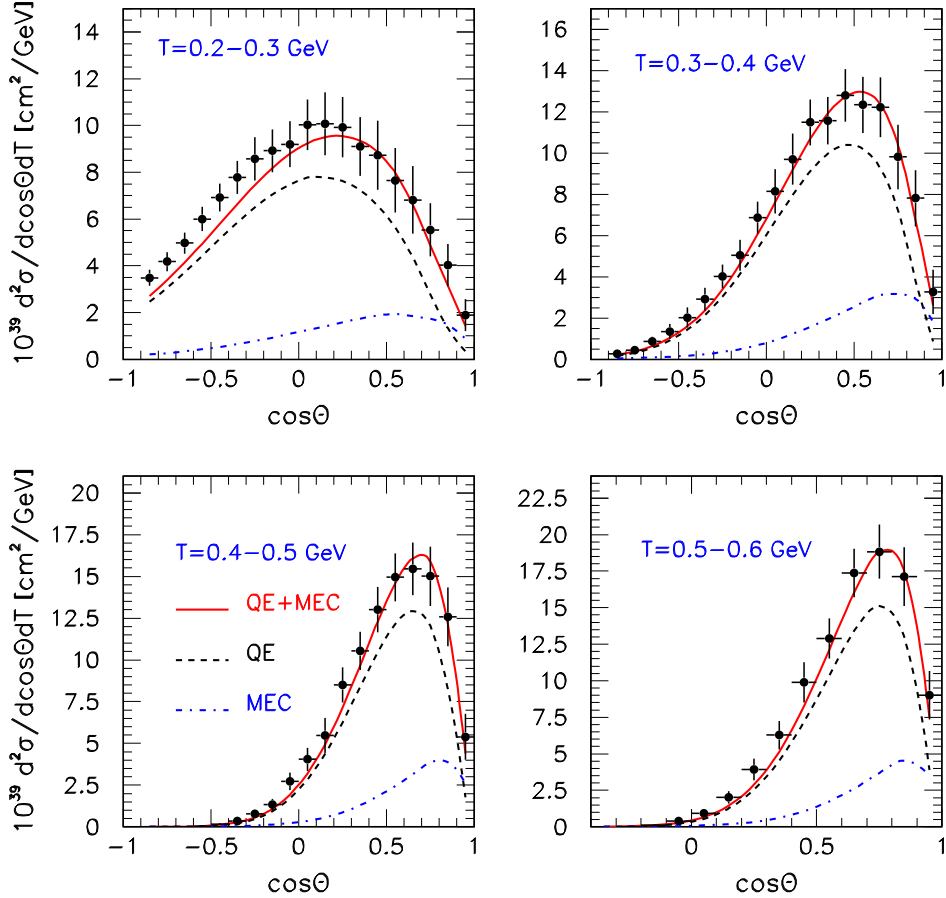}
  \end{center}
  \caption{\label{Fig5}Flux-integrated $d^2\sigma/dTd\cos\theta$ cross section
    per target neutron for the $\nu_{\mu}$ CCQE-like scattering as a function
    of $\cos\theta$ for the four muon kinetic energy bins: $T$(GeV)=(0.2-0.3),
    (0.3-0.4), (0.4-0.5), and (0.5 - 0.6). As shown in the key, cross sections
    were calculated within the RDWIA+MEC approach. The QE and
    $2p-2h$ MEC contributions are also presented. The MiniBooNE data are shown
    as points with the shape error only.} 
\end{figure*}
The following best fit
 $\chi^2$ and $M_A$ values are obtained: $\chi^2$/DOF=19/13 and
$M_A=1.17\pm 0.03$ GeV for the 1D fit, $\chi^2$/DOF=62/136 and
$M_A=1.24\pm 0.09$ GeV for the 2D fit, and $\chi^2$/DOF=111/150 and
$M_A=1.20\pm 0.06$ GeV for the 1D+2D main fit. Although there is a difference
between the best fit $M_A$ values, the errors from the fits cover this
difference.
\begin{figure*}
  \begin{center}
    \includegraphics[height=14cm,width=14cm]{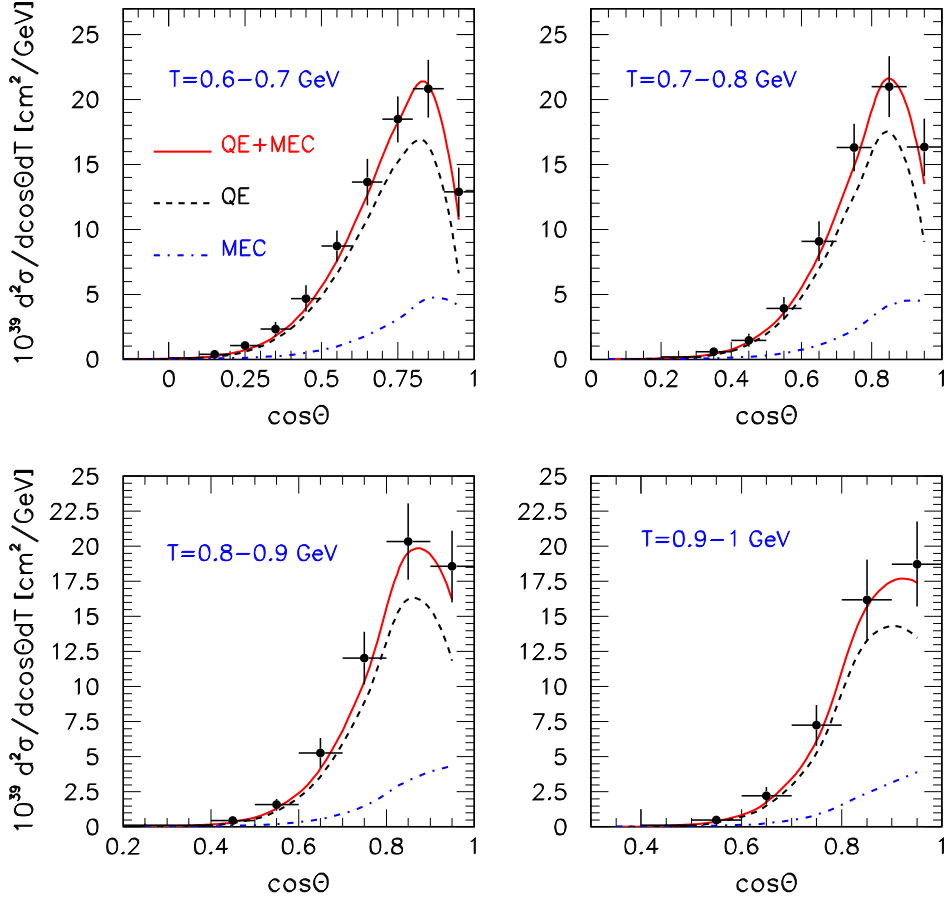}
  \end{center}
  \caption{\label{Fig6} Same as~Fig.~\ref{Fig5} but for the four muon kinetic
    energy bins: $T$(GeV)=(0.6-0.7), (0.7-0.8), (0.8-0.9), and (0.9-1).} 
\end{figure*}
Additionally, the value of $M_A=1.20\pm 0.06$ GeV from main fit is
in agreement within the errors with the best fit value of $M_A=1.15\pm 0.03$
GeV obtained from the main CCQE fit of the MiniBooNE and MINERvA data in
Refs.~\cite{Wilk, Wilk2}. The best fit $d\sigma/dQ^2$ distribution is compared
with the data in Fig.~\ref{Fig2}(a). The result of the dipole parametrization
of $F_A(Q^2)$ with $M_A=1.2$ GeV is shown in Fig.~\ref{Fig2}(b). There is an
overall agreement between the RDWIA+MEC $d\sigma/dQ^2$ cross section and the
data, but the model slightly overestimate the data in the range $0.08<Q^2<0.3$
(GeV/c)${}^2$.
In Figs.~\ref{Fig3}, \ref{Fig4}, \ref{Fig5},
\ref{Fig6}, and~\ref{Fig7} we show double differential cross sections
calculated with $M_A=1.2$ GeV.
The results for $d^2\sigma/dTd\cos\theta$ cross sections against the kinematic
energy of the muon are shown in Figs.~\ref{Fig3},~\ref{Fig4}.
\begin{figure*}
  \begin{center}
    \includegraphics[height=14cm,width=14cm]{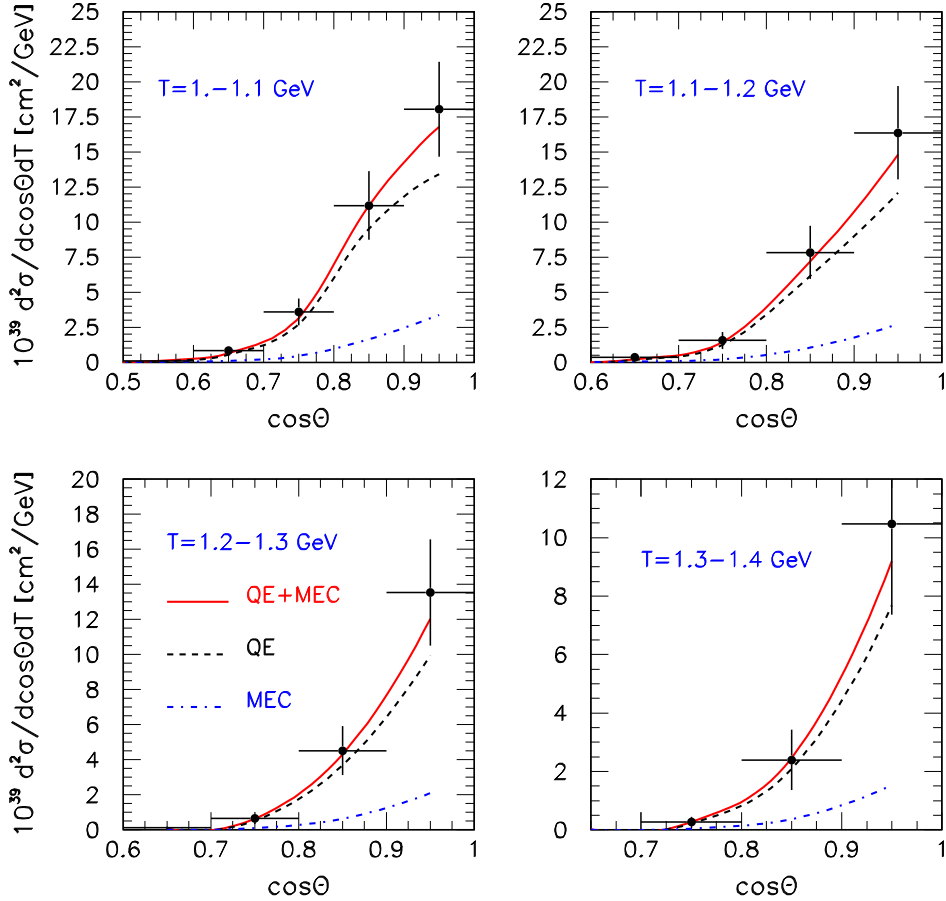}
  \end{center}
  \caption{\label{Fig7} Same as~Fig.~\ref{Fig5} but for the four muon kinetic
    energy bins: $T$(GeV)=(1.-1.1), (1.1-1.2), (1.2-1.3), and (1.3-1.4).} 
\end{figure*}
\begin{figure*}
  \begin{center}
    \includegraphics[height=14cm,width=14cm]{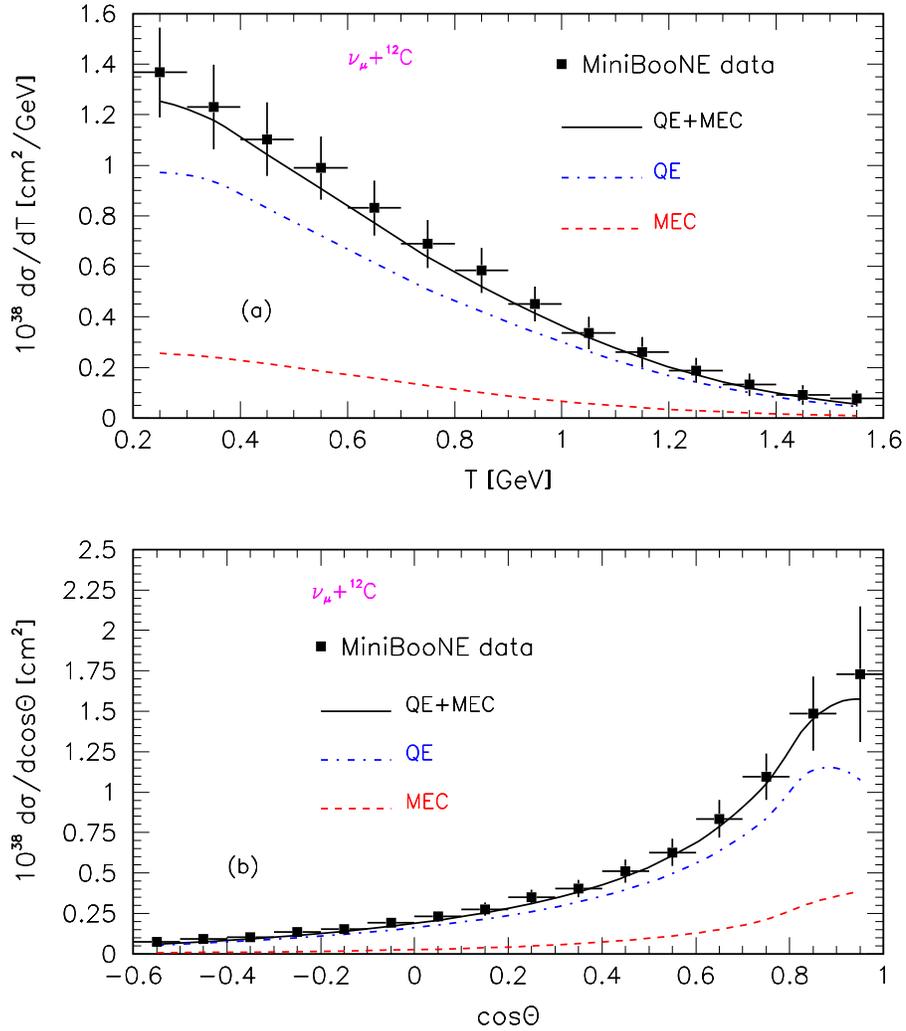}
  \end{center}
  \caption{\label{Fig8}Flux-integrated $d\sigma/dT$ cross section as a
    function of muon kinetic energy (upper panel) and
    $d\sigma/d\cos\theta$ cross section for $T>0.2$ GeV as a
    function of muon scattering angle (lower panel) for the $\nu_{\mu}$
    CCQE-like scattering per target neutron.
    As shown in the key, cross sections were calculated within the RDWIA+MEC.
    The RDWIA and $2p-2h$ MEC results are also presented separately. The
    MiniBooNE data are shown as points with the shape error only.} 
\end{figure*}
We present a large variety of kinematical situations where each panel refers to
results averaged over a particular muon angular bin. As observed, the model
tends to slightly underestimate the data for most forward angles, i.e.
$0.9 < \cos\theta < 1$. As the scattering angle increases, the RDWIA+MEC
prediction agrees well with the data. Results in Figs.~\ref{Fig3} and~
\ref{Fig4} clearly show that the $2p-2h$ MEC contributions are essential in
order to describe data. The contribution of these effects are comparable with
the genuine QE process, being of order 25\% and increasing up to 30\% at low
$Q^2$.
In Figs.~\ref{Fig5},~\ref{Fig6}, and~\ref{Fig7} we present the results
averaged over the muon kinetic energy bins as functions of the muon scattering
angle.
These graphs complement the previous ones, and show that the RDWIA+MEC model is
able to reproduce the data. There is a good agreement between the calculated
results and the data within experimental error. In the region
$0.2 < T < 0.3$ GeV and $-1 < \cos\theta < -0.2$ the model result is slightly
lower than the measured cross section, and the difference decreases with muon
energy.
In Fig.~\ref{Fig8} results are presented for the MiniBooNE flux-integrated
CCQE-like $d\sigma/dT$ differential cross section as a function of the muon
kinetic energy (upper panel) and $d\sigma/d\cos\theta$ cross section versus of
muon scattering angle (lower panel). The measured $d\sigma/dT$
($d\sigma/d\cos\theta$) cross section with the shape-only error was obtained
by summing the double-differential cross section over $\cos\theta$ bins
($T$ bins) presented in Tables VI and VII in Ref.~\cite{MiniB1}.
The integration over muon kinetic energy has been performed in the range
$0.2 < T < 2 $ GeV. As shown, the RDWIA+MEC model with $M=1.2$ GeV is capable
of reproducing the magnitude as well as the shape of the experimental cross
sections.

\begin{figure*}
  \begin{center}
    \includegraphics[height=10cm,width=12cm]{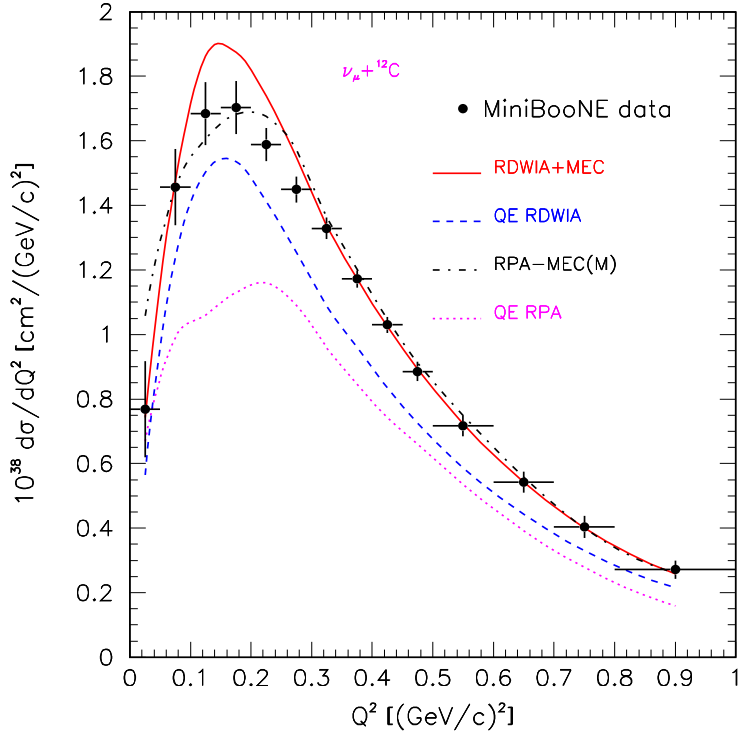}
  \end{center}
  \caption{\label{Fig9}Flux-integrated $d\sigma/dQ^2$ cross section per target
    neutron for the $\nu_{\mu}$ CCQE-like and CCQE processes as a function of
    $Q^2$. As shown in the key, cross sections were calculated
    within the RDWIA+MEC ($M_A=1.2$ GeV) and RPA-MEC~\cite{Martini2} models.
    The CCQE contributions calculated in the RDWIA and RPA approaches are also
    presented separately. The MiniBooNE data are shown as points with the
    shape-only error.} 
\end{figure*}

\begin{figure*}
  \begin{center}
    \includegraphics[height=14cm,width=14cm]{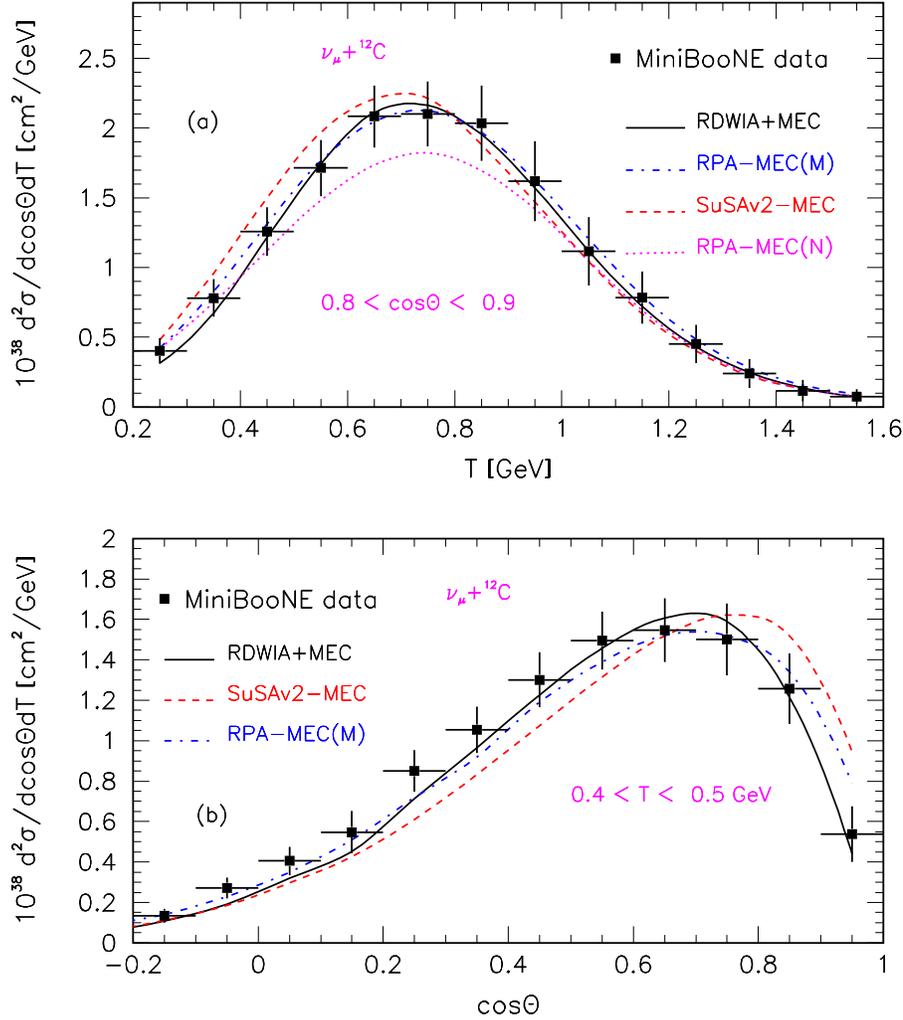}
  \end{center}
  \caption{\label{Fig10}Flux-integrated $d^2\sigma/dTd\cos\theta$ cross section
    per target neutron for the CCQE-like scattering. Upper panel: Cross
    sections calculated within the RDWIA+MEC (solid line), RPA-MEC
    (dashed-dotted line)~\cite{Martini2}, SuSAv2-MEC (dashed line), and
    RPA-MEC (dotted line)~\cite{Nieves2} models for $0.8<\cos\theta<0.9$ as
    functions of muon
    kinetic energy. Lower panel: Cross sections calculated in the RDWIA+MEC
    (solid line), SuSAv2-MEC (dashed line), and RPA-MEC (dashed-dotted line)
    ~\cite{Martini2} approaches for $0.4<T<0.5$ GeV as functions of muon
    scattering angle. The MiniBooNE data are shown as points with the
    shape-only error.} 
\end{figure*}

\begin{figure*}
  \begin{center}
    \includegraphics[height=14cm,width=14cm]{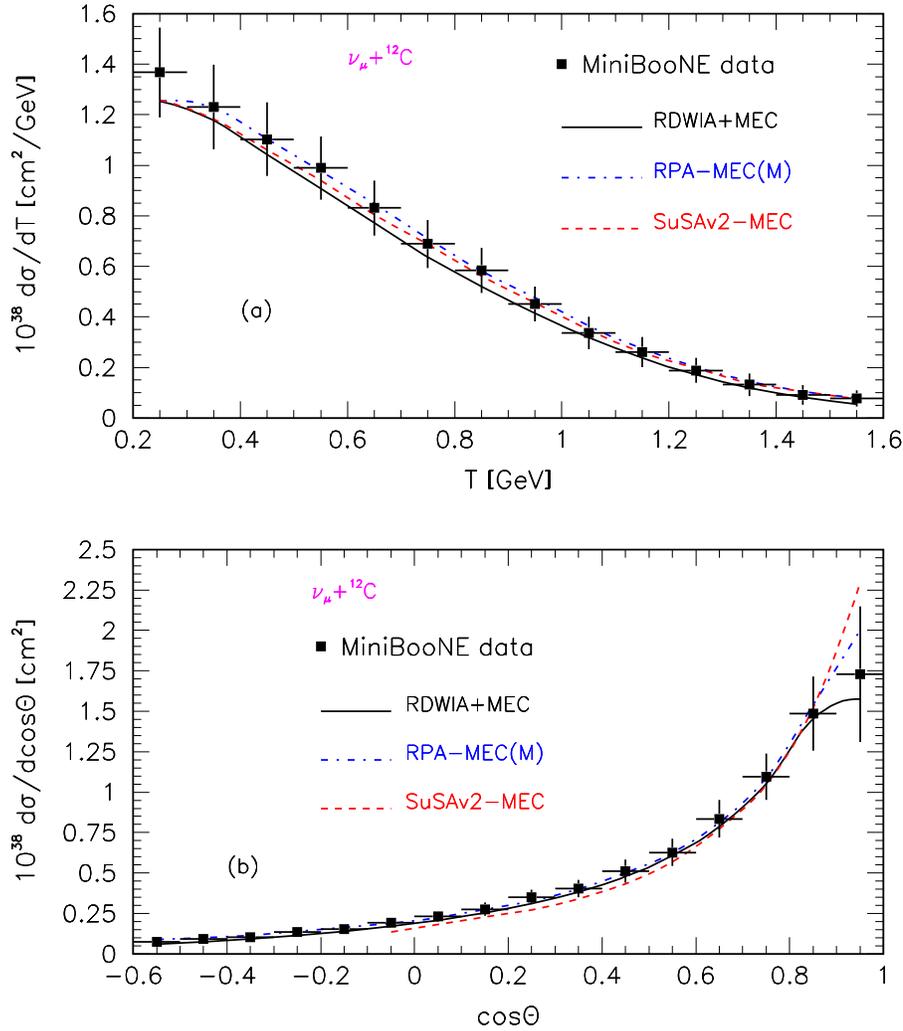}
  \end{center}
  \caption{\label{Fig11}Flux-integrated $d\sigma/dT$ cross section as a
    function of muon kinetic energy (upper panel) and
    $d\sigma/d\cos\theta$ cross section for $T>0.2$ GeV as a
    function of muon scattering angle (lower panel) for the CCQE-like
    scattering per target neutron. As shown in the key, cross sections
    were calculated within the RDWIA+MEC, SuSAv2-MEC, and RPA-MEC~
    \cite{Martini2} models. The MiniBooNE data are shown as points with the
    shape-only error.} 
\end{figure*}

In Figs.~\ref{Fig9}, \ref{Fig10}, and \ref{Fig11} the MiniBooNE neutrino
flux-averaged CCQE-like differential cross sections calculated within the
different approached are presented.
In Fig.~\ref{Fig9} we show the
$d\sigma/dQ^2$ cross sections as measured in Ref.~\cite{MiniB1} and as
calculated in the RDWIA+MEC and RPA-MEC~\cite{Martini2} models. Also shown are
CCQE cross sections obtained in these approaches.
From the figure one can
observe that these calculations describe well the experimental data at
$Q^2>0.3$ (GeV/c)${}^2$. The RDWIA+MEC model slightly overestimate the data in
the range $0.08<Q^2<0.3$ (GeV/c)${}^2$ and in the case of the RPA+MEC approach
a tendency to overestimate the data appears at low $Q^2<0.06$ (GeV/c)${}^2$.
In the range of $Q^2<0.3$ (GeV/c)${}^2$, which is affected by
RPA quenching, the CCQE cross sections calculated in the RDWIA with $M_A=1.2$
GeV is $\approx 30$\% larger than those obtained in Ref.~\cite{Martini2} and the
discrepancy decreases with $Q^2$ up to 12\% at $Q^2\approx 0.9$ (GeV/c)${}^2$.
In Fig.~\ref{Fig10} we show the double-differential $d^2\sigma/dTd\cos\theta$
cross sections calculated in the RDWIA+MEC, SuSAv2-MEC~\cite{Megias2}, and
RPA-MEC~\cite{Nieves2, Martini2} approaches.
For the sake of illustration in
Fig.~\ref{Fig10}(a) the results are given for $0.8<\cos\theta<0.9$
as functions of the muon kinetic energy. As one can observe, the results of the
RDWIA+MEC, SuSAv2-MEC, and RPA-MEC~\cite{Martini2} models are in agreement with
data. In the case of the RPA-MEC result~\cite{Nieves2} a tendency to
underestimate the MiniBooNE data appears.
In Fig.~\ref{Fig10}(b) the results are given
for muon kinetic energy bin $0.4<T<0.5$ GeV as functions of the muon scattering
angle. As shown, the results obtained within the RDWIA+MEC and RPA-MEC~
\cite{Martini2} models agree well with data. On the other hand, a difference
between the SuSAv2-MEC result and data is observed.
The flux-averaged differential cross sections $d\sigma/dT$ and
$d\sigma/d\cos\theta$ (for $T>0.2$ GeV), calculated in the RDWIA+MEC,
SuSAv2-MEC, and RPA-MEC~\cite{Martini2} approaches are presented in
Fig.~\ref{Fig11}, which shows $d\sigma/dT$ as a function of muon kinetic energy
and $d\sigma/d\cos\theta$ as a function of the muon scattering angle.
\begin{figure*}
  \begin{center}
    \includegraphics[height=14cm,width=14cm]{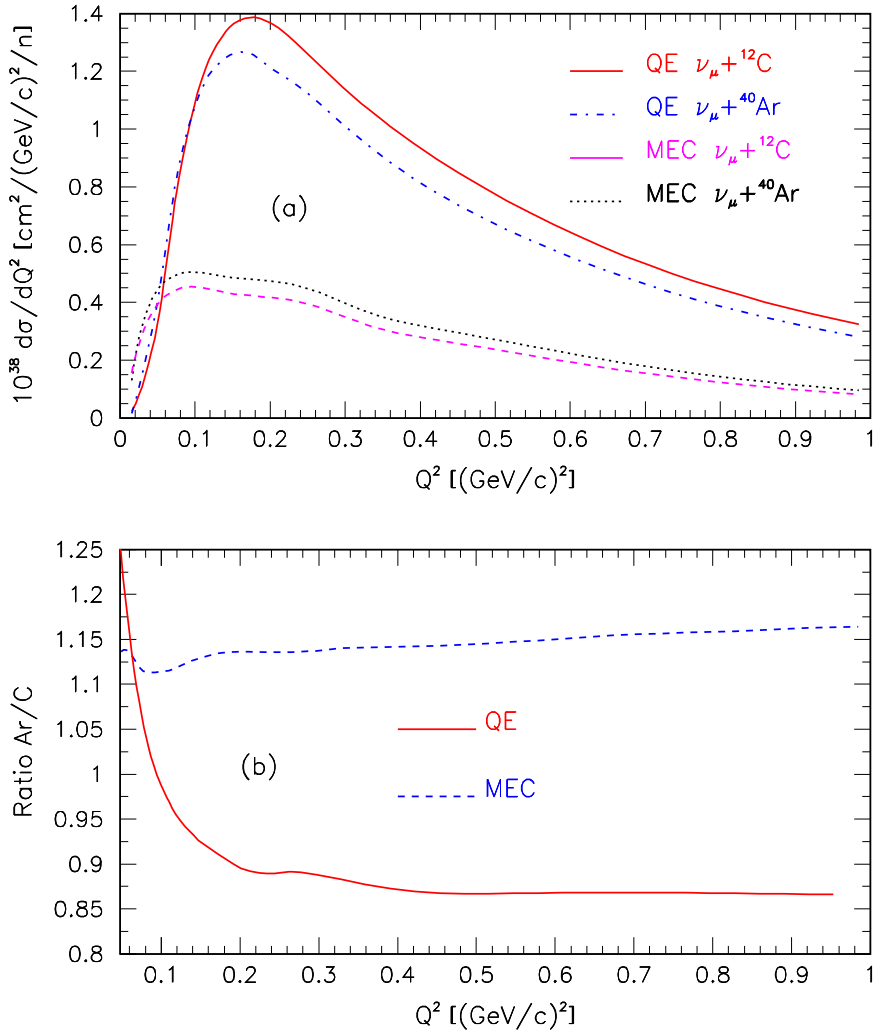}
  \end{center}
  \caption{\label{Fig12}NOvA flux-integrated $d\sigma/dQ^2$ cross sections per
    target neutron for the $\nu_{\mu}$ CCQE and $2p-2h$ MEC scattering
    (upper panel) and ratio $R_{QE}$ and $R_{MEC}$ (lower panel) as functions of
    $Q^2$. As shown in the key, cross sections were calculated for neutrino
    scattering on ${}^{12}$C and ${}^{40}$Ar.}
\end{figure*}
Also shown are the MiniBooNE measured cross sections with the shape-only error.
There is a good agreement within the errors between the calculated results and
data.

\subsection {Calculation of neutrino CCQE-like differential cross sections at 
 energies of the NOvA experiment}

Within the RDWIA+MEC approach with $M_A=1.2$ GeV, which was successfully tested
against the MiniBooNE neutrino data, we estimated the neutrino CCQE-like
flux-integrated differential cross sections at energies available at the NOvA
experiment~\cite{NOvA1, NOvA2}.
\begin{figure*}
  \begin{center}
    \includegraphics[height=10cm,width=14cm]{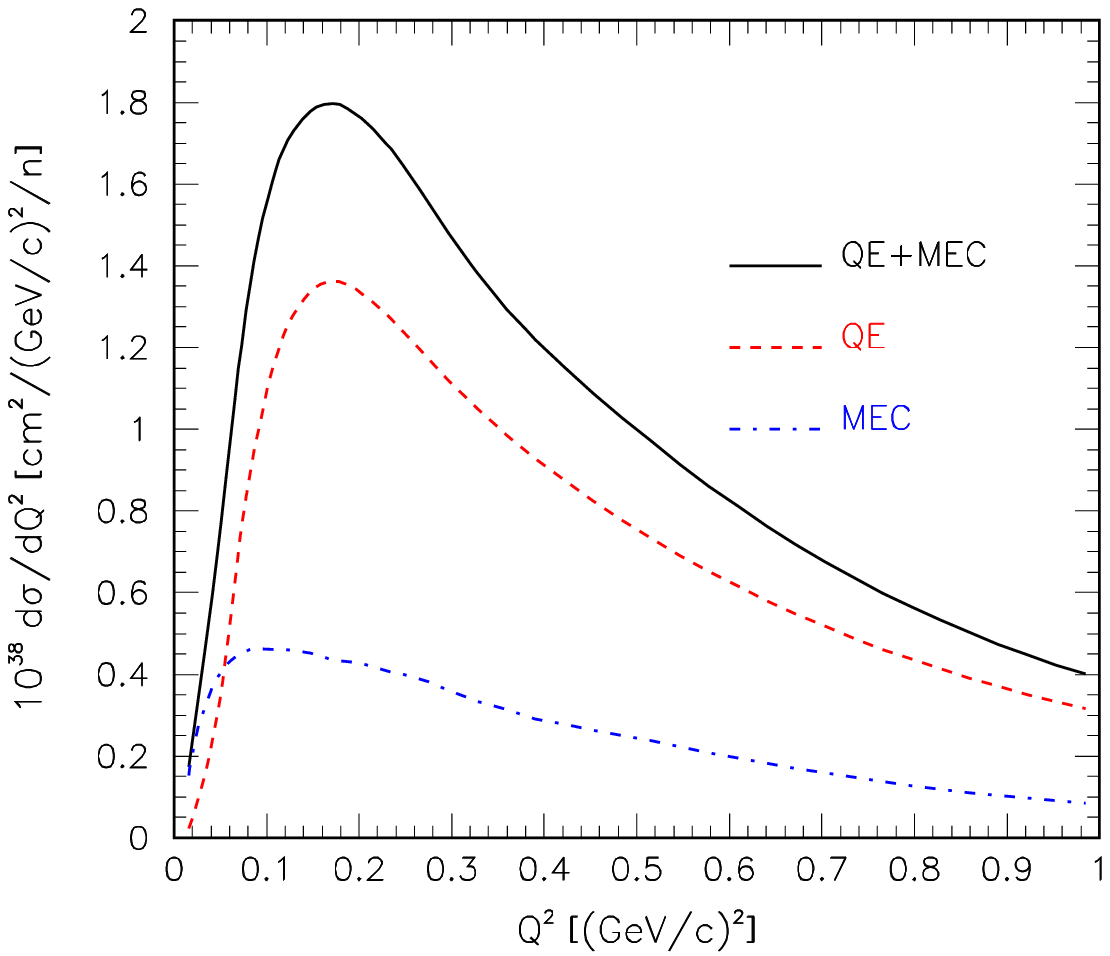}
  \end{center}
  \caption{\label{Fig13}NOvA flux-integrated $d\sigma/dQ^2$ cross sections per
    target neutron for the $\nu_{\mu}$ CCQE-like scattering (solid line). The
    CCQE (dashed line) and $2p-2h$ MEC (dashed-dotted line) results are also
    presented separately. The cross sections are shown as functions of $Q^2$.}
\end{figure*}
The NOvA detectors are situated 14
mrad off the neutrino main injector (NuMI) beam axis, so they expose a
relatively narrow band $\sim 0.5 - 5$ GeV of neutrino energies, centered at 2
GeV~\cite{Leo2}. This flux is used in the calculation of the
NOvA flux-integrated differential cross sections.

In the fiducial region of the NOvA near detector (ND) the liquid scintillator
(CH$_2$) comprises 63\% of the detector mass. Mass weight for this detector
component is as follows: ${}^{12}$C - 66.8\%, ${}^{35}$Cl - 16.4\%, ${}^{1}$H -
10.5\%, ${}^{48}$Ti - 3.3\%, ${}^{16}$O - 2.6\%, and others - 0.4\%~
\cite{Xuebing}. So, the ND is dominanted by carbon, chlorine, and hydrogen. We
assumed that the NOVA CCQE-like scattering sample consists of two processes:
scattering on ${}^{12}$C and ${}^{35}$Cl. The mass weight of carbon $\alpha_c$
and chlorine $\alpha_{Cl}$ was re-scaled for neutrino scattering as
$\alpha_C=0.806$, $\alpha_{Cl}=0.194$, and $\alpha_c + \alpha_{Cl}=1$. In
Ref.~\cite{BAV5} we calculated within the RDWIA the CCQE differential cross
sections for (anti)neutrino scattering on ${}^{40}$Ar. The difference between  
the nuclear structures of ${}^{40}$Ar and ${}^{35}$Cl is not significant.
\begin{figure*}
  \begin{center}
    \includegraphics[height=14cm,width=14cm]{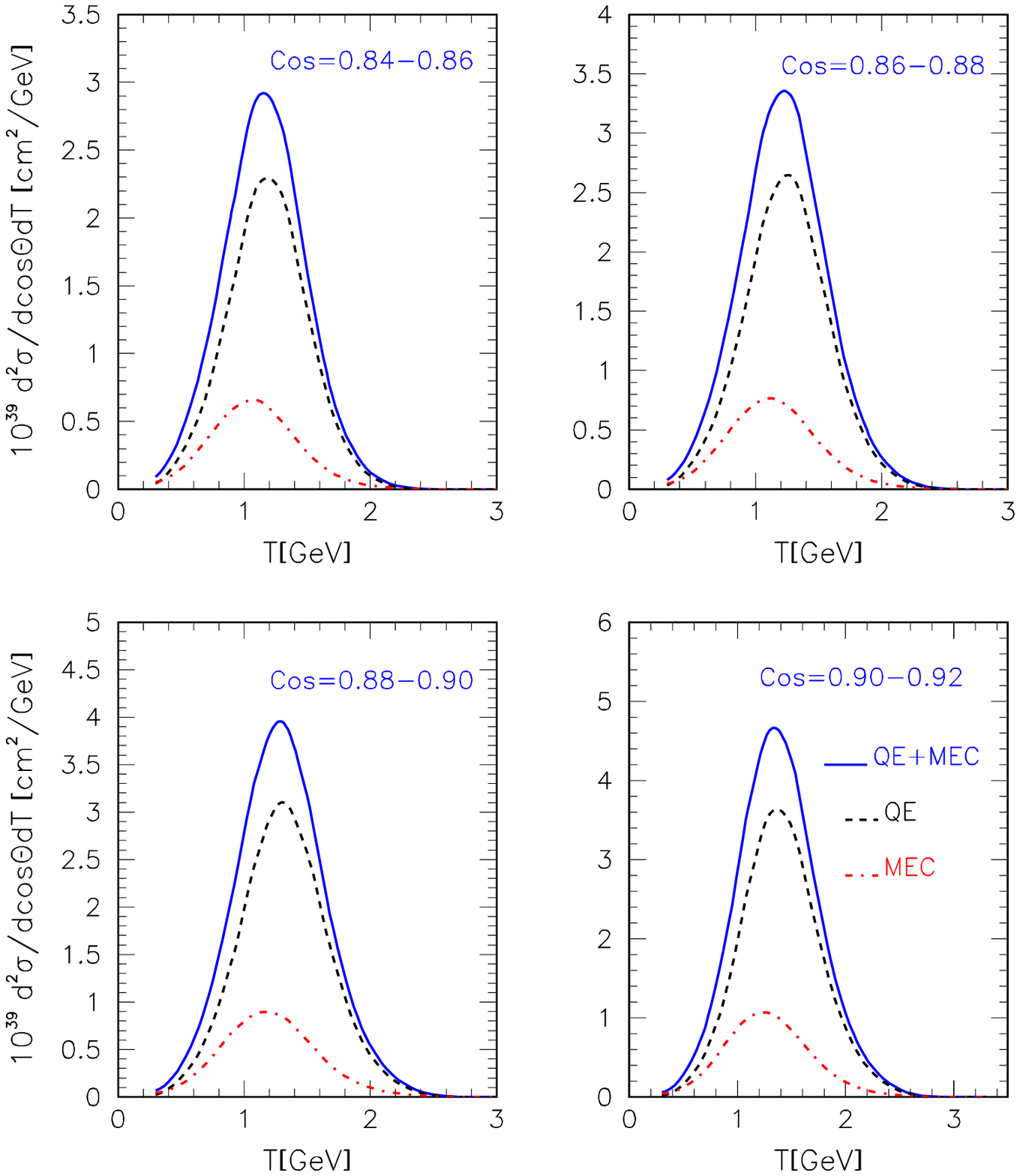}
  \end{center}
  \caption{\label{Fig14}NOvA flux-integrated $d^2\sigma/d\cos\theta\ dT$ cross
    section per target neutron for the $\nu_{\mu}$ CCQE-like scattering as a
    function of muon kinetic energy for the four muon scattering angle bins:
    $\cos\theta$=(0.84-0.86), (0.86-0.88), (0.88-0.90), and (0.90-0.92). As
    shown in the key, cross section was calculated within the RDWIA+MEC. 
   The CCQE and $2p-2h$ MEC contributions are also shown separately.}
\end{figure*}
\begin{figure*}
  \begin{center}
    \includegraphics[height=14cm,width=14cm]{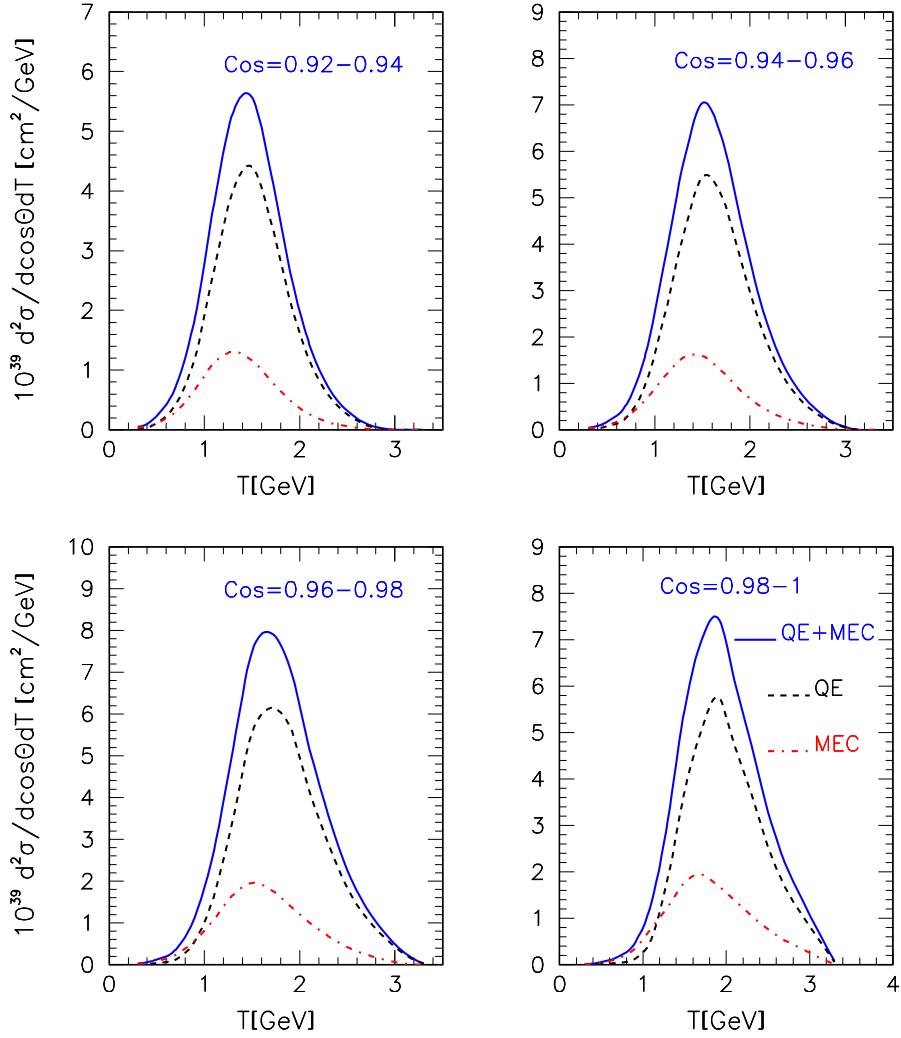}
  \end{center}
  \caption{\label{Fig15}Same as Fig.~\ref{Fig14} but for muon scattering
    angle bins: $\cos\theta$=(0.92-0.94), (0.94-0.96), (0.96-0.98), and
    (0.98-1).}
\end{figure*}
\begin{figure*}
  \begin{center}
    \includegraphics[height=14cm,width=14cm]{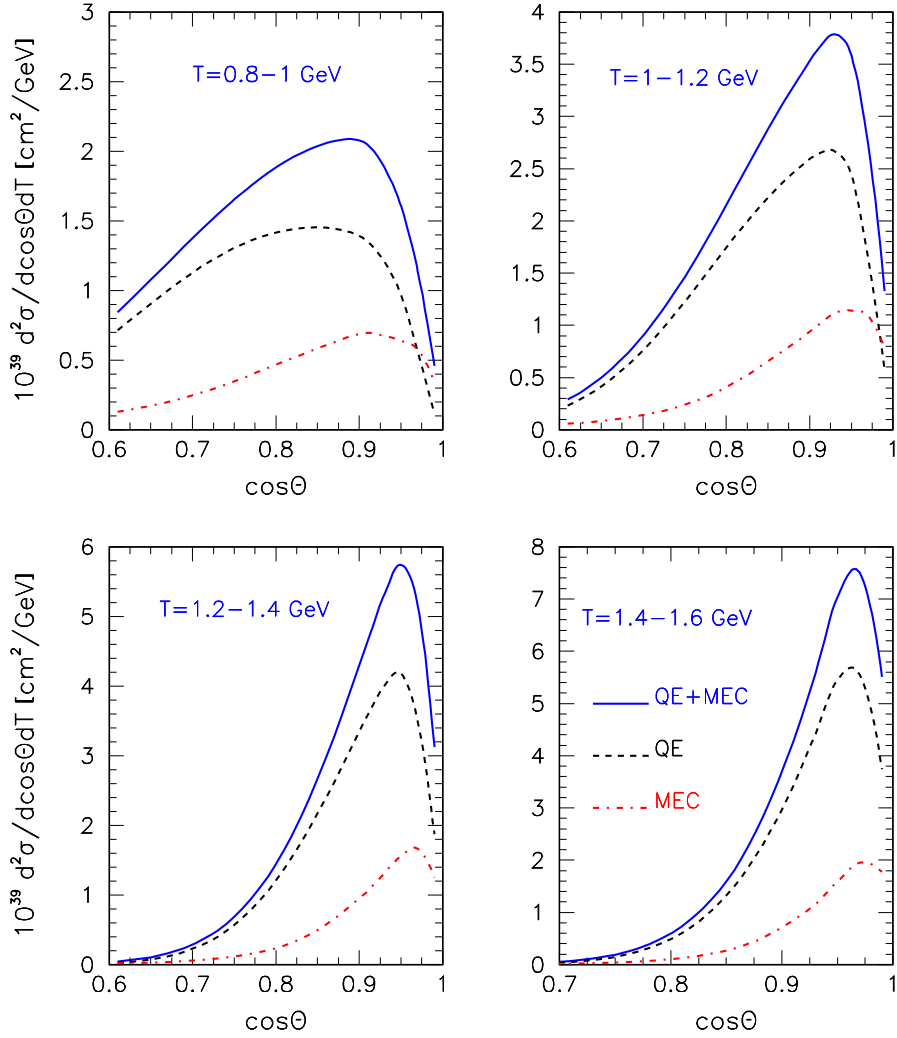}
  \end{center}
  \caption{\label{Fig16}NOvA flux-integrated $d^2\sigma/d\cos\theta dT$ cross
    section per target neutron for the $\nu_{\mu}$ CCQE-like scattering as a
    function of muon kinetic energy for the four muon kinetic energy bins:
    $T$(GeV)=(0.8-1), (1-1.2), (1.2-1.4), and (1.4-1.6). As shown in key,
    cross section was calculated within the RDWIA+MEC. The CCQE and $2p-2h$
    MEC contributions are also presented separately.}
\end{figure*}
\begin{figure*}
  \begin{center}
    \includegraphics[height=14cm,width=14cm]{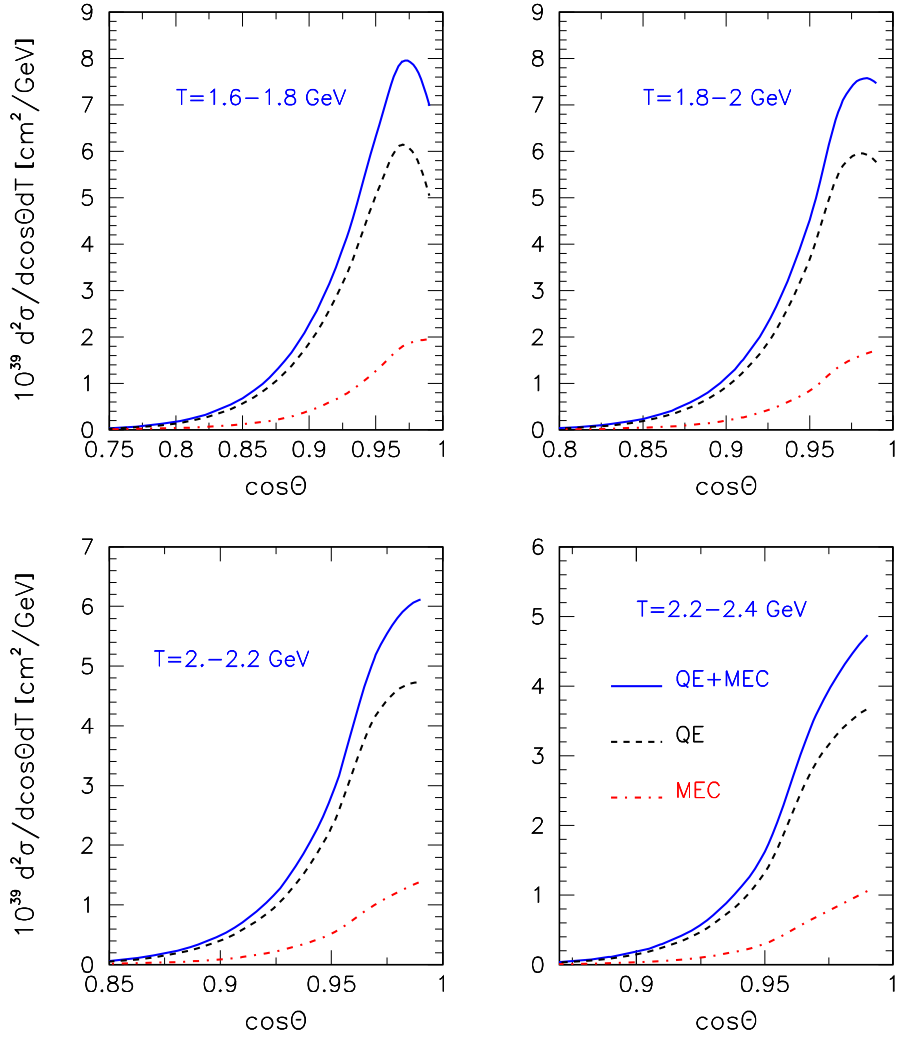}
  \end{center}
  \caption{\label{Fig17}Same as Fig.~\ref{Fig16} but for muon kinetic energy
    bins: $T$(GeV)=(1.6-1.8), (1.8-2.0), (2.0-2.2), and (2.2-2.4).}
\end{figure*}
\begin{figure*}
  \begin{center} 
    \includegraphics[height=12cm,width=12cm]{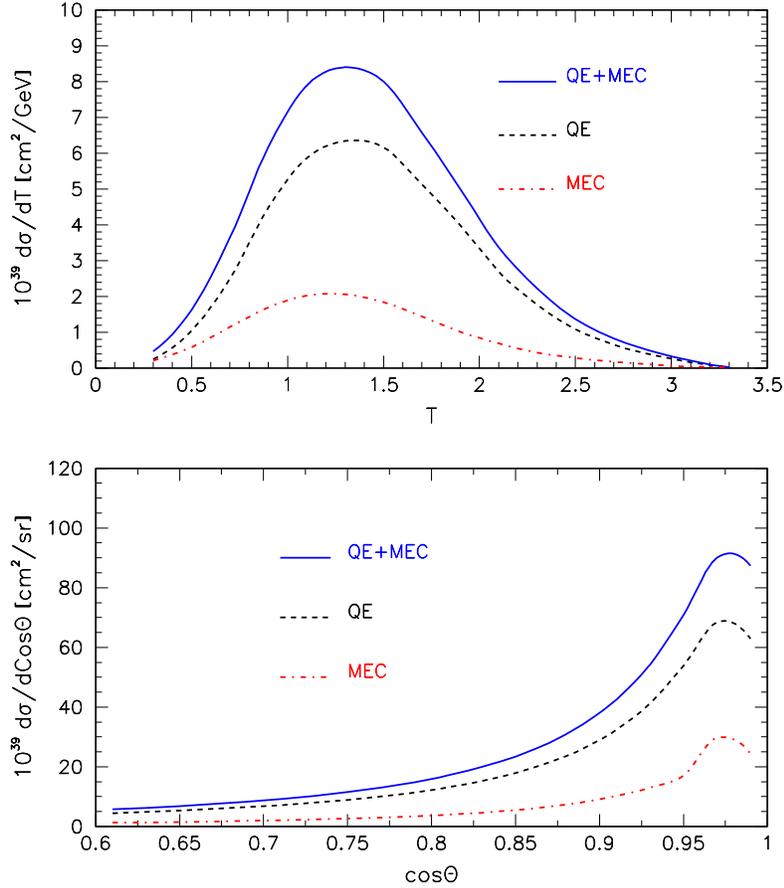}
  \end{center}
  \caption{\label{Fig18}NOvA flux-integrated $d\sigma/dT$ cross
    sections for $0.6<\cos\theta<1$ as a function of muon kinetic energy
    (upper panel) and $d\sigma/d\cos\theta$ cross section for $0.2<T<3.5$ GeV
    as a function of muon scattering angle (lower panel) for the $\nu_{\mu}$
    CCQE-like scattering per target neutron. As shown in the key, cross
    sections were calculated within the RDWIA+MEC approach. The CCQE and
    $2p-2h$ MEC contributions are also shown.}
\end{figure*}
\begin{figure*}
  \begin{center}
    \includegraphics[height=10cm,width=11cm]{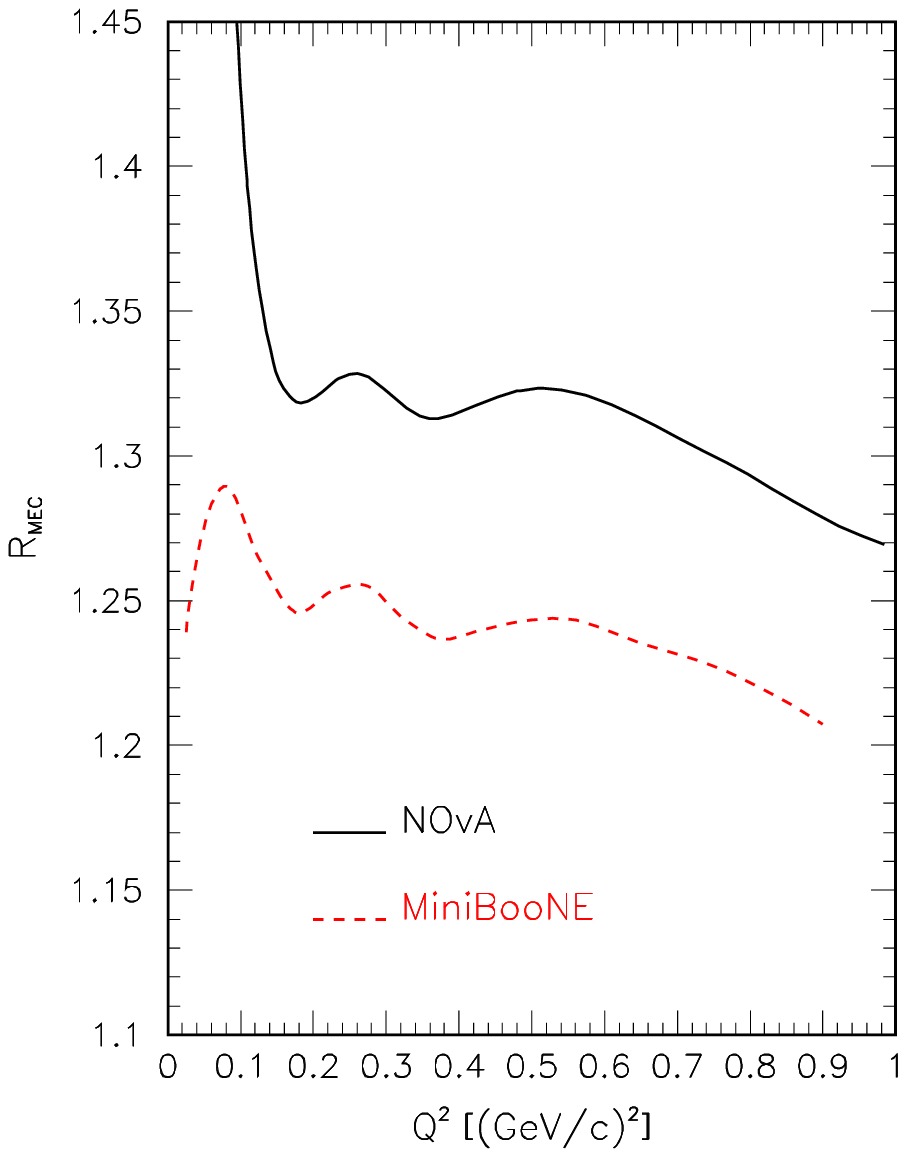}
  \end{center}
  \caption{\label{Fig19}The ratio $R_{MEC}$ as a function of $Q^2$ for the
    NOvA and MiniBooNE flux-integrated $d\sigma/dQ^2$ cross sections.}
\end{figure*}
Therefore the NOvA CCQE-like differential cross sections were estimated for
neutrino scattering on carbon and argon, herewith the $2p-2p$ MEC contributions
for ${}^{40}$Ar was calculated using the parameterizations for ${}^{12}$C
re-scaled for argon according to~\cite{Amaro5}. Then the NOvA neutrino
scattering cross section per target neutron, {\it i.e.} the cross sections averaged
over the ND mass weight, can be expressed as
$\sigma_{MIX}=\alpha_C \sigma_C + \alpha_{Cl}\sigma_{Ar}$,
where $\sigma_c(\sigma_{Ar})$ is the cross section of neutrino scattering on
${}^{12}$C(${}^{40}$Ar).

Fig.~\ref{Fig12} contains the RDWIA+MEC model predictions corresponding to the
NOvA flux-integrated $d\sigma/dQ^2$ cross sections per target neutron for
neutrino scattering on ${}^{12}$C and ${}^{40}$Ar. The cross sections were
calculated for the CCQE and $2p-2h$ MEC processes in the region $T>0.3$ GeV
and $0.3<\cos\theta<1$. Also shown are the ratios
$R_{QE}=(d\sigma/dQ^2)^{Ar}_{QE}/(d\sigma/dQ^2)^C_{QE}$
 and $R_{MEC}=(d\sigma/dQ^2)^{Ar}_{MEC}/(d\sigma/dQ^2)^C_{MEC}$, where 
 $(d\sigma/dQ^2)^{Ar}_{QE}[(d\sigma/dQ^2)^C_{QE}]$ and
 $(d\sigma/dQ^2)^{Ar}_{MEC}[(d\sigma/dQ^2)^C_{MEC}]$ are the CCQE and $2p-2h$ MEC
 cross sections per target neutron for neutrino scattering on ${}^{40}$Ar
 (${}^{12}$C), respectively. The figure clearly shows that the ratio $R_{QE}$
 reduces with $Q^2$ from 1.2 at $Q^2\approx 0.04$ (GeV/c)${}^2$ up to  0.87 at
 $Q^2\approx 1$ (GeV/c)${}^2$. On the other hand the ratio $R_{MEC}$ increases
 slowly with $Q^2$ from  1.1 at $Q^2\approx 0.1$ (GeV/c)${}^2$ up to  1.17
 at $Q^2\approx 1$ (GeV/c)${}^2$. Thus, the RDWIA+MEC model predicts that the
 CCQE differential cross section per target neutron reduces and $2p-2h$ MEC
 contribution increases with the mass number of the target.

 The results in Fig.~\ref{Fig13} correspond to the flux-integrated NOvA
 $(d\sigma/dQ^2)_{MIX}$ cross section per target neutron of the CCQE-like
 neutrino scattering. Also shown are the contributions of the CCQE and $2p-2h$
 MEC processes.
 The ratio $R=(d\sigma/dQ^2)_{MIX}/(d\sigma/dQ^2)_C$ is about 0.98 in
 the range $0.1<Q^2<1$ (GeV/c)${}^2$, {\it i.e.} the NOvA CCQE-like cross section per
 target nucleon for neutrino scattering in the NOvA ND is, practically, the same
 as one for scattering on carbon.

 In Figs.~\ref{Fig14} and~\ref{Fig15} we present the flux-integrated CCQE-like
 double differential cross sections per neutron predicted for the NOvA
 experiment. The graphs are plotted against the muon kinetic energy and each
 panel corresponds to a bin in the muon scattering angle. The double
 differential cross sections averaged over muon kinetic energy bins are shown in
 Figs.~\ref{Fig16} and \ref{Fig17}. In these figures we show the separate
 contributions of the genuine QE and $2p-2h$ MEC processes. The NOvA
 flux-integrated CCQE-like $\nu_{\mu}$ cross sections per target neutron
 $d\sigma/dT$ as a function of the muon kinetic energy and $d\sigma/d\cos\theta$
 versus muon scattering angle are presented in Fig.~\ref{Fig18}. The pure QE
 and $2p-2h$ MEC results are also shown separately. Integration of the double
 differential cross section over muon scattering angle has been performed in
 the range $0.6 < \cos\theta < 1$, for calculation of the $d\sigma/dT$ cross
 section and over muon kinetic energy in the range $0.2<T<3.5$ GeV, for
 evaluation of the $d\sigma/d\cos\theta$ cross section.

 The ratio of the neutrino flux-integrated differential CCQE-like 
 $(d\sigma/dQ^2)_{QE+MEC}$ cross section to the genuine QE $(d\sigma/dQ^2)_{QE}$ 
 one, $R_{MEC}=(d\sigma/dQ^2)_{QE+MEC}/(d\sigma/dQ^2)_{QE}$, calculated for the
 MiniBooNE and NOvA experiments are shown in Fig.~\ref{Fig19}. As observed, in
 the NOvA case the contribution of the $2p-2h$ MEC is about 8\% higher than
 in the MiniBooNE experiment. This can be connected with the NOvA neutrino flux
 that is centered at neutrino energy $\approx 2$ GeV, whereas the MiniBooNE
 neutrino flux has maximum at the energy $\approx 0.7$ GeV. As was discussed 
 earlier, the $\Delta$-peak is the main contribution to the pion production
 cross section which increases with neutrino energy in the range up to 3 GeV.
 
\section{Conclusions}

In this article we analyzed the MiniBooNE neutrino data within the RDWIA+MEC
model. This model has been validated in the vector sector by describing the
set of inclusive electron scattering ${}^{12}$C data. We performed a shape-only
fit of the RDWIA+MEC approach to the data with only the nucleon axial mass, as
variable model parameter. A best fit value of $M_A=1.2 \pm 0.06$ GeV was
obtained. This value is in agreement within the errors with the best fit value
of $M_A=1.15 \pm 0.03$ GeV obtained from the CCQE main fit of the MiniBooNE
and MINERvA data~\cite{Wilk, Wilk2}. We also extracted the values of the axial
form factor $F_A(Q^2)$ as a function of $Q^2$, using the measured neutrino
flux-integrated $d\sigma/dQ^2$ cross section. There is a good overall agreement
within the experimental uncertainties between the extracted values of
$F_A(Q^2)$ and the dipole parametrization with the value of $M_A=1.2$ GeV. We
obtained that in the MiniBooNE experiment the $2p-2h$ channel is large,
contributing about 25\% depending on kinematics, and it is essential to
describe a great amount of experimental data. One can also notice that our
calculations are in agreement with other theoretical results which are
compatible with MiniBooNE data.   

Using the RDWIA+MEC model with  $M_A=1.2$ GeV we estimated the differential
flux-integrated CCQE-like cross sections for neutrino scattering in the NOvA
near detector. We show that these cross crosections are coincidence with ones
for neutrino scattering on carbon. The $2p-2h$ MEC contributions in the NOvA
energy range are about of 30-35\%, i.e. $\sim 8$ \% larger than in the
MiniBooNE experiment. So, the measurements of the CCQE-like differential cross
sections in the NOvA experiment are necessary  in order to make precision
determinations of neutrino oscillation parameters.

\section*{Acknowledgments}

The authors greatly acknowledge J. Amaro and G. Megias, for fruitful 
discussions  and for putting in our disposal the codes for calculation of the 
MEC's electroweak response functions that were used in this work. We specially
thank A. Habig and J. Samoilova for a critical reading of the manuscript.

%


\end{document}